\documentclass[aps,prd,superscriptaddress,nofootinbib,amsfonts,amssymb,amsmath,notitlepage,twocolumn]{revtex4-1}

\usepackage[dvipdfmx]{graphicx}
\usepackage{amsmath}
\usepackage{amssymb} \allowdisplaybreaks[0]
\usepackage{float}
\usepackage[colorlinks=true]{hyperref}
\usepackage{makecell}
\usepackage{multirow}
\usepackage{slashed}
\usepackage{wrapfig}
\usepackage{bm}
\usepackage{color}
\usepackage{xcolor}
\usepackage[nameinlink]{cleveref}
\hypersetup{
	colorlinks,
	linkcolor={blue!75!black},
	citecolor={blue!75!black},
	urlcolor={blue!75!black}
}

\newcommand{\Slash}[1]{{\ooalign{\hfil/\hfil\crcr$#1$}}} 
\newcommand{\bvec}[1]{\bm{#1}}
\newcommand{\p}{\partial}
\newcommand{\nn}{\nonumber\\}
\newcommand{\df}{\text{d}}
\newcommand{\Tr}{{\rm Tr}\,}
\newcommand{\pmat}[1]{\begin{pmatrix}#1\end{pmatrix}}

\graphicspath{{./figs/}}
\newbox{\ORCIDicon}
\sbox{\ORCIDicon}{\large
                  \includegraphics[width=0.8em]{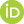}}

\begin{document}

\title{
QCD chiral phase diagram from weak functional renormalization group
}

\author{Yuepeng \surname{Guan}\,\href{https://orcid.org/0009-0007-8571-0931}{\usebox{\ORCIDicon}}}
%\email{e}
\affiliation{Center for Theoretical Physics and College of Physics, Jilin University, Changchun 130012, China}

\author{Masatoshi \surname{Yamada}\,\href{https://orcid.org/0000-0002-1013-8631}{\usebox{\ORCIDicon}}}
%\email{m.yamada@kwansei.ac.jp}
\affiliation{Department of Physics and Astronomy, Kwansei Gakuin University, Sanda, Hyogo, 669-1330, Japan}

\begin{abstract}
We investigate the QCD chiral phase transition at finite temperature and finite baryon density using the functional Renormalization Group (fRG).
While conventional fRG studies often employ techniques such as dynamical bosonization to regularize divergences, we instead pursue the weak solution of the fRG equations which allows for non-analytic behavior in the flow to compute the pure fermionic potential $V_k(\psi,\bar{\psi})$ within the local potential approximation. 
This approach enables us to explore the effects of purely quark-level fluctuations on dynamical chiral symmetry breaking without introducing any auxiliary bosonic fields.
Based on this framework, we present the resulting chiral phase diagram as a function of temperature and baryon chemical potential.
\end{abstract}
\maketitle
% \tableofcontents

%%%%%%%%%%%%%%%%%%%%%%%%%%%%%%%%%%%%%%%%%%%%%%%%%%%%%%%%%%%%%%%%%%%%%%
\section{Introduction}

Quantum Chromodynamics (QCD), as the asymptotically free gauge theory of the strong interaction, has achieved remarkable success in explaining the confinement and collective dynamics of quarks and gluons. 
Its predictive power underpins modern nuclear and particle physics, from the structure of hadrons to the formation of quark-gluon plasma (QGP) in relativistic heavy-ion collisions.

Progress in elucidating the QCD phase diagram has revealed a rich landscape of nontrivial phenomena. At high temperature and low baryon chemical potential, lattice QCD simulations and heavy-ion experiments confirm the crossover transition from hadronic matter to a deconfined QGP~\cite{Jacob:1982qk, Kapusta:2003qyx}. 
Conversely, in the large baryon chemical potential region, 
it is expected that a critical endpoint exists and first-order transitions take place at large baryon chemical potential, while the phase structure remains still enigmatic because Monte-Carlo methods falter due to the sign problem.
Recent advances in functional methods, such as the Dyson-Schwinger equations (DSEs)~\cite{Welzbacher:2014pea, Gao:2024ggp, Gao:2020qsj, Lu:2023mkn, Lu:2025cls} and the functional Renormalization Group (fRG)~\cite{Cyrol:2017ewj, Cyrol:2017qkl, Fu:2019hdw}, have provided complementary insights into chiral symmetry restoration and color superconductivity. 

In QCD, especially in the chiral limit, i.e. the vanishing current quark masses, the chiral phase transition is dominated by collective mesonic fluctuations. 
Naive pure quark-based fRG frameworks struggle to resolve the chiral phase transition due to the divergence of the four-fermion interaction, which reflects the massless nature of the collective mode.
To address the symmetry-broken phase within the fRG framework, we need to employ techniques which appropriately capture the fluctuations of emergent degrees of freedom.
Dynamical bosonization~\cite{Gies:2001nw, Gies:2002hq, Pawlowski:2005xe, Floerchinger:2009uf, Braun:2014ata, Mitter:2014wpa, Cyrol:2017qkl, Cyrol:2017ewj, Alkofer:2018guy,Denz:2019ogb, Fu:2019hdw,Goertz:2024dnz} addresses the chiral phase transition by explicitly introducing bosonic fields (mesons) as emergent degrees of freedom, enabling a more natural description of symmetry-breaking phenomena.
Alternatively, recent works also encounter the full momentum dependence in the four-quark vertex function~\cite{Fu:2024ysj, Fu:2025hcm}, overcoming limitations of early bosonization schemes that relied on local approximations. These advances allow us to precisely evaluate mesonic spectra and critical fluctuations near the QCD critical endpoint.

Another way to access the symmetry-broken phase is through the ``weak fRG" formulation~\cite{Aoki:2014ola,Aoki:2013gda,Aoki:2017rjl}.
The strategy of the method is to utilize the ``weak form'' of the fRG equations, which relaxes continuity requirements, allowing solutions with nonanalytic behavior (e.g., discontinuities in derivatives of the effective potential), which is critical for capturing spontaneous symmetry breaking.
This approach allows us to explore chiral symmetry breaking solely through fermionic fluctuations.

\begin{figure}
    \centering
    \includegraphics[width=0.9\linewidth]{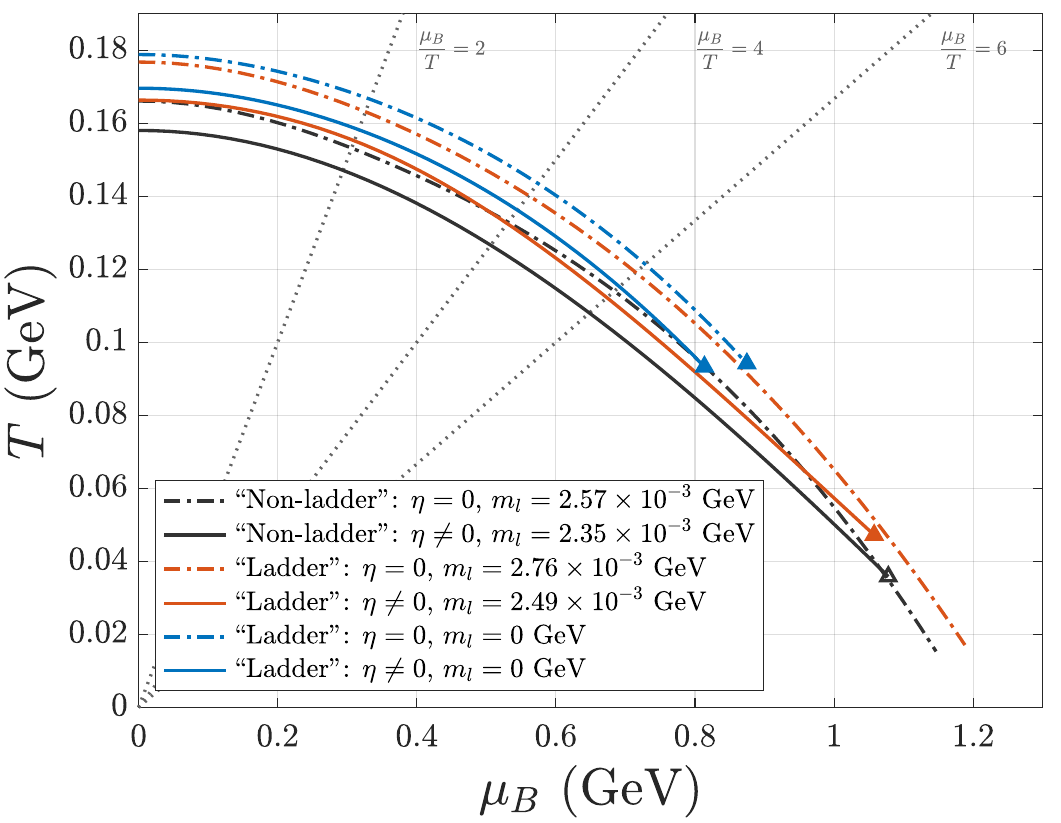}
    \caption{QCD phase diagrams in terms of the phase boundary ($T_c$-lines) of second-order (chiral limit) or crossover (physical point) within the parameter setups in \Cref{tab:parameterFixing}.
    The positions of the critical end point (CEP) in each case are marked by triangle markers.}
    \label{fig:TogetherPhaseDiagram}
\end{figure}

In this paper, we solve the flow equation of the fermionic potential $V_k(\psi,\bar{\psi})$ at finite temperature and quark chemical potential through the search for the weak solution of the fRG equation.
In particular, we evaluate the dynamical mass and obtain the QCD chiral phase diagram on the $(\mu_B,T)$ plane.
To derive the flow equation for $V_k(\psi,\bar{\psi})$, we employ the ladder approximation and its non-ladder extensions, considering both cases---with and without the anomalous dimension of the quark field---in the chiral limit as well as for finite quark masses. 
The flow equation for $V_k(\psi,\bar{\psi})$ is then reformulated into a weak form of a differential equation. 
For the running gauge coupling, we adopt input data obtained from an independent fRG analysis~\cite{Gao:2021wun}.
The resulting phase diagram is shown in \Cref{fig:TogetherPhaseDiagram}.

This paper is organized as follows.
In \Cref{sec:functionalRenormalizationGroupSetup}, we summarize the necessary elements, such as the regulator choice and approximation scheme, in setting up the flow equation of $V_k(\psi,\bar{\psi})$.
In \Cref{sec:WeakSolutionAndMethodOfCharacteristics}, we summarize the logic flow of the weak fRG equation and the method of characteristics.
We present our results after briefly introducing our numerical setups, in \Cref{sec:results}.
\Cref{sec:summary} is devoted to the summary and conclusions.
Detailed derivations are provided in \Cref{app:WetterichEquationWithSharpRegulator,app:DeriveFlowEq}.

%%%%%%%%%%%%%%%%%%%%%%%%%%%%%%%%%%%%%%%%%%%%%%%%%%%%%%%%%%%%%%%%%%%%%%
\section{Functional renormalization group setup}
\label{sec:functionalRenormalizationGroupSetup}

We start with the bare QCD action at finite temperature and finite chemical potential in the Euclidean spacetime % at the ultraviolet (UV) scale $\Lambda$
\begin{align}
    S_{\rm bare}[\Phi] & = \int_x \, \biggl[ \frac{1}{4} F^a_{\mu\nu} F^a_{\mu\nu} + \frac{1}{2 \xi} (\partial_\mu A_\mu^a)^2 \nonumber\\
    &\quad \quad + \bar{\psi} (\Slash{\partial} + i g_s \Slash{A} - m_l - \mu_q \gamma_4 ) \psi \biggl],
    \label{eq:bareQCDAction}
\end{align}
where the spacetime integral is given by $\int_x = \int_0^{1/T} \df x_4 \int \df^3 x$ with temperature $T$ and the field content reads
\begin{align}
    \Phi = (A^a_\mu, \psi^T, \bar{\psi} )^T.
    \label{eq:field component}
\end{align}
In \cref{eq:bareQCDAction}, we denote $\Slash{A} = A^a_\mu \gamma_\mu T^a$ with $\gamma_\mu$ being the gamma matrices in the Euclidean spacetime and $T^a$ being the matrix representations of the $SU(3)$ gauge group.
The gauge field strength is given by $F^a_{\mu\nu} = \partial_\mu A^a_\nu - \partial_\nu A^a_\mu + g_s f^{abc} A^b_\mu A^c_\nu$. 
The fermionic field $\psi = (u^T,d^T)^T$ includes the light two flavors of quarks, and we adopt the isospin symmetry for the current quark mass $m_u = m_d \equiv m_l$.
The light quark chemical potential $\mu_q$ can be replaced by the baryonic chemical potential $\mu_B$ as $\mu_q = \mu_B/3$.
The second term in \cref{eq:bareQCDAction} is the gauge-fixing term with the gauge-fixing parameter $\xi$; the corresponding ghost action is omitted for brevity.

To study the phase structure from the QCD action \labelcref{eq:bareQCDAction} non-perturbatively, we employ the Wetterich equation~\cite{Wetterich:1992yh} (see also \cite{Ellwanger:1993mw, Morris:1993qb})
\begin{align}
    \label{eq:WetterichEq}
    \partial_t \Gamma_k[\Phi] &= \frac{1}{2} \operatorname{STr} \left[ \left( \Gamma_k^{(2)} + \mathcal{R}_k \right)^{-1} \partial_t \mathcal{R}_k \right],
\end{align}
where $\Gamma_k$ is the one-particle irreducible (1PI) averaged effective action and $t = \log(k/\Lambda)$ is the dimensionless RG scale with a reference scale $\Lambda$.
The Wetterich equation \labelcref{eq:WetterichEq} describes the evolution of $\Gamma_k$ with respect to changing $k$ from the initial action $\Gamma_{k=\Lambda\to \infty}=S_\text{bare}$. 
Here $\Gamma_k^{(2)}$ is the inverse full two-point function (the Hessian) which is obtained by the second-order functional derivative with respect to the fields \labelcref{eq:field component}
\begin{align}
    \Gamma_{k,I,J}^{(2)}(q,p) = \frac{\overset{\rightarrow}{\delta}}{\delta \Phi^T_I(-p)} \Gamma_k  \frac{\overset{\leftarrow}{\delta}}{\delta \Phi_J(q)},
    \label{eq:defineHessian}
\end{align}
where the indices $I$ and $J$ stand for the internal degrees of freedom in the fields, and STr is the functional supertrace acting on all spaces in which the fields are defined.
The coarse-graining process in the formulation of the Wetterich equation \labelcref{eq:WetterichEq} is controlled by the regulator function $\mathcal R_k$.

\subsection{Truncated effective action}

In this work, we employ the modified local potential approximation (LPA') scheme to truncate the effective action $\Gamma_k$ in a solvable way.
Performing the derivative expansion to the effective average action and keeping the lowest order, we make an ansatz for the effective action
\begin{align}
    &\Gamma_k[\Phi] 
     = \int_x \, \biggl[ \frac{Z_k^A}{4} F^a_{\mu\nu} F^a_{\mu\nu} + \frac{Z_k^A}{2 \xi} \int_x (\partial_\mu A_\mu^a)^2 \nn 
    &\quad \quad  + Z_k^\psi \bar{\psi}(\Slash{\partial} + i g_s \Slash{A} - \mu_q \gamma_4 )\psi - V_k(\psi,\bar{\psi}) \biggl].
    \label{eq:IREA}
\end{align}
Here $V_k(\psi,\bar\psi)$ is the fermionic potential given as polynomials of gauge-invariant bilinear operators $\bar\psi \Gamma\psi$ where $\Gamma$ represents the spinor basis such as ${\bf 1}_\text{spinor}$, $\gamma_\mu$, etc, and $Z_k^\psi$ and $Z_k^A$ are the field renormalization factors for the quark fields and for the gluon field, respectively.
In \cref{eq:IREA}, we have neglected the gluonic (Polyakov loop) potential, which implies that we do not consider the confinement effect.
Also, we do not introduce the thermal splitting of the Lorentzian tensorial structures as an approximation, which means that we introduce the unified wave function renormalization factor for the temporal and spatial components of the kinetic terms for both gluon and fermion.

For $\mathcal R_{k}$, we adopt regulator functions for the three-dimensional momentum, i.e.,
\begin{align}
&\mathcal R_k^\psi({p}) = Z_k^\psi i{\Slash{{p}}}r_k^\psi(|\bvec{p}|/k),
\label{eq:Rfermion}
\\[1ex]
&(\mathcal R_k^A(p))_{\mu\nu} = Z_k^A p^2 \Pi^{1/\xi}_{\mu\nu} \,r_k^\psi(|\bvec{p}|/k),
\label{eq:Rhgauge}
\end{align}
with the sharp-cutoff regulator function in momentum space
\begin{align}
    r_k^A = r_k^\psi = \frac{1}{\theta(|\bvec{p}|/k - 1)} - 1.
\label{eq:dimlessregulator}
\end{align}
Here, $\Pi^{1/\xi}_{\mu\nu}$ is the tensor structure of the two-point function for the gluon field at the tree level (see \cref{eq:propAndParaProjection}) and the color indices were suppressed in \cref{eq:Rhgauge}.
For this choice, the flow equation \labelcref{eq:WetterichEq} reads
\begin{align}
    \partial_t \Gamma_k[\Phi] = - \frac{1}{2} \int_{p,\text{shell}}\operatorname{str}\log \Gamma_k^{(2)}.
    \label{eq:flowinsharpcutoff}
\end{align}
Here and hereafter we introduce a shorthand notation for the momentum integral
\begin{align}
 \int_{p,\text{shell}}\equiv T\sum_{n = -\infty}^{\infty} \int \frac{\df^3 \bvec{p}}{(2\pi)^3} {k\delta\left( |\bvec{p}| - k \right)}.
\end{align}
Note here that the regulators \labelcref{eq:Rfermion,eq:Rhgauge} allow us to sum over all Matsubara modes in the temporal direction and to perform the momentum integral analytically.
The derivation of \cref{eq:flowinsharpcutoff} and the general structure of the flow equation are summarized in \Cref{app:WetterichEquationWithSharpRegulator}.

\subsection{Flow of the fermionic potential and approximations}
\label{sec:flowOfVkMainText}

We discuss the derivation of the flow equation for the fermionic potential $V_k(\psi,\bar\psi)$.  
In this subsection, we present the necessary components to clarify the approximation scheme employed in the present work.  
The complete derivation of the flow equation for the fermionic potential is provided in \Cref{app:DeriveFlowEq}.

At each order in the fermionic field expansion, the potential $V_k(\psi,\bar\psi)$ contains various non-trivial tensorial structures.  
In the chiral limit and ignoring the $U(1)_A$ anomaly, chiral symmetry is exact at the level of the action and can only be broken spontaneously (at the vacuum level), so the operators in the effective action must be chirally invariant.
For instance, the fermionic effective potential includes four-fermion interactions that are invariant under flavor transformations:
\begin{align}
 V_k(\psi,\bar\psi) \ni \left( \bar{\psi} \lambda^I\psi \right)^2 + \left( \bar{\psi} \lambda^I i \gamma_5 \psi \right)^2 + \cdots,
\end{align}
where $\lambda^I$ denotes the generators of the flavor symmetry group, with $\lambda^0 = \frac{1}{\sqrt{N_f}}\mathbf{1}_{\text{flavor}}$.

Here, we adopt a further approximation by projecting the flow onto the subspace that contains only flavor and Dirac singlet channels, i.e., the subspace involving only the scalar bilinear $\sigma = \bar{\psi}\psi$.  
In this case, the fermionic effective potential reduces to $V_k(\psi, \bar\psi ) \equiv V(\sigma;t)$.  
Consequently, the potential respects a discrete chiral symmetry under $\psi \to \psi$ and $\bar\psi \to -\bar\psi$, rather than a continuous one.  
Thus, Nambu--Goldstone modes (pseudoscalar fluctuations) are not included.  
Neglecting contributions from other tensorial structures, such as $(\bar\psi\gamma^\mu\psi)^2$, may lead to the so-called Fierz ambiguity (see, e.g., \cite{Jaeckel:2002rm,Braun:2017srn,Braun:2018bik,Braun:2019aow}).  
Nevertheless, our aim in this paper is to investigate the impact of non-ladder effects on the phase diagram.  
Therefore, we argue that this treatment introduces only quantitative differences in the final results and does not affect the qualitative physics under discussion.

In the chiral limit, the discrete chiral symmetry forbids odd-power terms of $\sigma$ in the fermionic potential. 
However, going beyond the chiral limit due to the presence of current quark masses, the chiral symmetry is explicitly broken, and the fermionic potential then contains odd powers of $\sigma$.
Thus, the reduced fermionic potential formally reads
\begin{align}
V(\sigma;t) = \sum_{n=1} \frac{G^{(n)}_k}{n!} \sigma^n = \sum_{n=1} \frac{G^{(n)}_k}{n!} (\bar\psi\psi)^n .
    \label{eq:scalarBilinearChannelOfVkExpansion}
\end{align}
Note here that $G_k^{(2)}$ corresponds to the four-fermi coupling for $(\bar\psi\psi)^2$.

When evaluating the Hessian for the fermionic part, we retain only the leading term in the large-$N$ expansion, neglecting the diagonal elements in the fermionic sector. In this approximation, the flow equation for the fermionic effective potential depends solely on its first derivative with respect to $\sigma$, i.e., $\partial_\sigma V(\sigma;t)$; see \cref{eq:secondFermionDerivativeLargeN}.
After performing the functional derivatives, we take the following external field backgrounds
\begin{align}
    A_\mu^a = 0, \quad \psi(x) = e^{i \pi T x_0} \psi, \quad \bar{\psi}(x) = e^{-i \pi T x_0} \bar{\psi},
\end{align}
where we take into account the lowest mode of the fermion thermal fluctuation according to its statistical feature.
Combining all those setups, the Hessian is obtained within the LPA' framework, see \cref{eq:LPAPrimeHessian} for the explicit expression.

After inserting the Hessian into the flow equation \labelcref{eq:flowinsharpcutoff}, and performing the field renormalization as
\begin{align}
    A_\mu^a \rightarrow A^a_\mu /\sqrt{Z^{A}_k} , 
    \quad 
    \psi \rightarrow \psi/\sqrt{Z^{\psi}_k},
    \quad
    \bar\psi \rightarrow \bar\psi/\sqrt{Z^{\psi}_k},
\end{align}
the flow equation is then reduced to
\begin{align}
    &\partial_t V(\sigma;t) =  \eta_\psi \sigma \partial_\sigma V(\sigma;t) \nonumber\\
    &\quad - \frac{1}{2} \int_{p, \rm shell} \operatorname{tr}\biggl[ \log{S^{-1}(p^-_\psi)} + \log{(S^{(T)}(p^+_\psi))^{-1}} \biggl] \nonumber\\
    &\quad + \frac{1}{2} \int_{p,\rm shell} \operatorname{tr}^\prime\log \biggl\{ \delta^{ab} \delta_{\mu\nu} + \mathcal{A}^{ab}_{\mu\nu} + \mathcal{B}^{ab}_{\mu\nu} \biggl\},
    \label{eq:flowOfVkInGeneralMaintext}
\end{align}
where the explicit definition of $S^{-1}(p^-_\psi)$, $(S^{(T)}(p^+_\psi))^{-1}$, $\mathcal{A}^{ab}_{\mu\nu}$ and $\mathcal{B}^{ab}_{\mu\nu}$ can be found in \cref{eq:LPAPrimeHessian,eq:inverseQuarkPropagators,eq:defineAandBtensors1,eq:defineAandBtensors}, respectively, and the prime on the trace indicates that the trace acts only on the Lorentz and adjoint color spaces.
%The fermionic zero-point energy subtraction mentioned in \cref{eq:subtructionTerms} has been omitted here.
In \cref{eq:flowOfVkInGeneralMaintext}, $\eta_\psi = - \partial_t \log Z^\psi_k$ denotes the quark anomalous dimension, which is derived in \Cref{App:QuarkAnomalousDimension}.
Since we do not introduce the thermal split of the fermionic kinetic energy, which is the same setup utilized in Ref.~\cite{Fu:2019hdw}, we project the flow of the quark fluctuation onto the tensorial structure $i\bvec{\gamma}\cdot\bvec{p}$ to represent the flow of the quark wave function $Z^\psi_k$, and thus read off the quark anomalous dimension as
\begin{align}
\eta_\psi = -\frac{1}{8N_fN_c}\lim_{\bvec p\to0}\frac{\p^2}{\p |\bvec p|^2}\Tr\left(i\bvec{\gamma}\cdot\bvec{p}\frac{\delta^2\Gamma_k}{\delta\bar\psi(-p)\delta\psi(p)} \right).
\end{align}
See \cref{eq:quarkAnomalousDimension} in \Cref{App:QuarkAnomalousDimension} for the explicit form of the quark anomalous dimension.

For the purpose of solving the fermionic potential $V_k$ with respect to the RG scale $k$ and the external field background $\sigma$, we need a closed form of the flow equation.
To this end, we introduce the so-called ``ladder'' and ``non-ladder'' approximation schemes~\cite{Aoki:2000dh,Aoki:2012mj} evaluating the trace of the second term in \cref{eq:flowOfVkInGeneralMaintext}. See also \cite{Aoki:1999dw}.

\subsubsection{``Ladder'' case}

\begin{figure}
    \centering
    \includegraphics[width=0.9\linewidth]{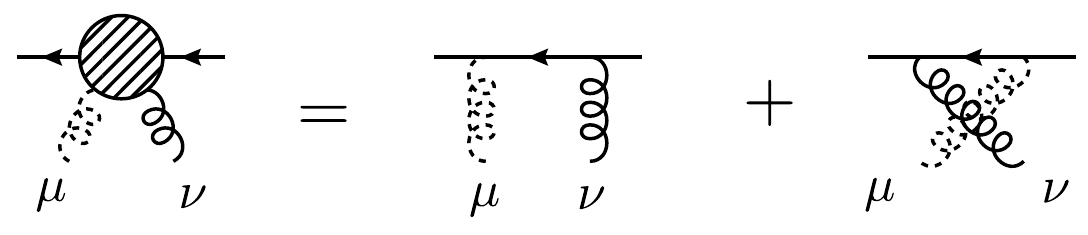}
    \caption{Sketch of the mixed quark-gluon vertex $\mathcal{A} + \mathcal{B}$. The solid line represents a fermion, while the curly lines correspond to gluons.}
    \label{fig:MixedVertex}
\end{figure}

The strategy of evaluating the $\operatorname{tr}^\prime\log (1+\mathcal{A}+\mathcal{B})$ is to expand the expression in terms of $(\mathcal{A}+\mathcal{B})^n$ corresponding to the $n$-th vertex expansion with respect to $\sigma$.
Diagrammatically, $\mathcal{A}$ and $\mathcal{B}$ denote two types of effective vertices of the gluon propagator dressing; we call $(\mathcal{A}+\mathcal{B})$ the mixed quark-gluon vertex.
This is depicted diagrammatically in \Cref{fig:MixedVertex}.

When expanding $(\mathcal{A}+\mathcal{B})^n$, the terms $\mathcal{A}^l\mathcal{B}^m$ ($l+m=n$) represent the different constructions of the gluonic loop with fermionic background fields.
Truncating or approximating this combination is to drop some types of loops in the flow.

\begin{figure}
    \centering
    \includegraphics[width=0.9\linewidth]{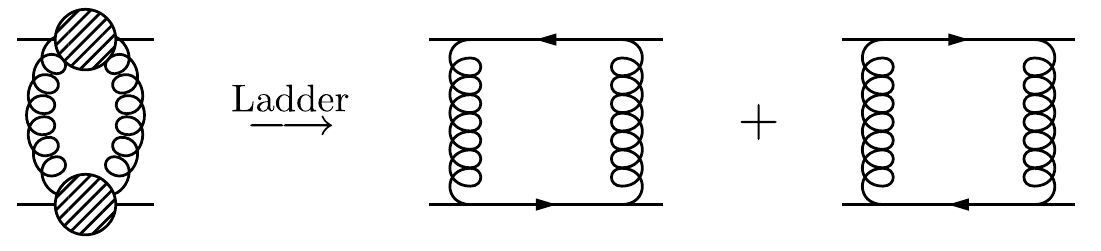}
    \caption{Sketch of $(\mathcal{A}+\mathcal{B})^2 \to \mathcal A^2+ \mathcal B^2$ under the ladder approximation \labelcref{eq:ladderAproxMainText}.}
    \label{fig:LadderApprox}
\end{figure}

The simplest approximation scheme is to keep only the loops with forward or reverse ordering of the color generator, i.e.,
\begin{align}
    (\mathcal{A}+\mathcal{B})^n \rightarrow \mathcal{A}^n + \mathcal{B}^n.
    \label{eq:ladderAproxMainText}
\end{align}
As an example, we schematically show the diagrams that are taken into account for the $4$-point function, i.e., $(\mathcal{A}+\mathcal{B})^2$-term, in \Cref{fig:LadderApprox}.
From this figure, this approximation scheme is called the ``ladder'' approximation.

After taking this approximation and the general Fierz transformation toward the scalar channel, the vertex expansion is then able to be resummed in a closed form easily.
This results in
\begin{align}
    &\partial_t V(\sigma ; t) =  \eta_\psi \sigma \partial_\sigma V(\sigma;t) \nn
    &\quad- T\sum_n  \frac{k^3}{2\pi^2} \log\left( \omega_{\psi,n}^2 + ( \sqrt{k^2+ \mathcal M^2} - \mu_q )^2 \right) \nn
    &\quad- T\sum_n  \frac{k^3}{2\pi^2} \log\left( \omega_{\psi,n}^2 + ( \sqrt{k^2+\mathcal M^2} + \mu_q )^2 \right),
    \label{eq:ladderFlowMainText}
\end{align}
where the dressed mass of quarks is defined as
\begin{align}
\mathcal M \equiv \partial_\sigma V(\sigma;t) + C_2 g_s^2 \frac{3 + \xi}{4 ( k^2 + \omega_{A,n}^2)} \sigma,
    \label{eq: B definition}
\end{align}
with Matsubara frequencies $\omega_{\psi,n} = (2n+1)\pi T$ and $\omega_{A,n} = 2n \pi T$, where $C_2=\sum_{a=1}^{N_c^2-1}T^a T^a$ is the Casimir operator of the fundamental representation of $SU(N_c)$.

\subsubsection{``Non-ladder'' case}

\begin{figure}
    \centering
    \includegraphics[width=0.2\linewidth]{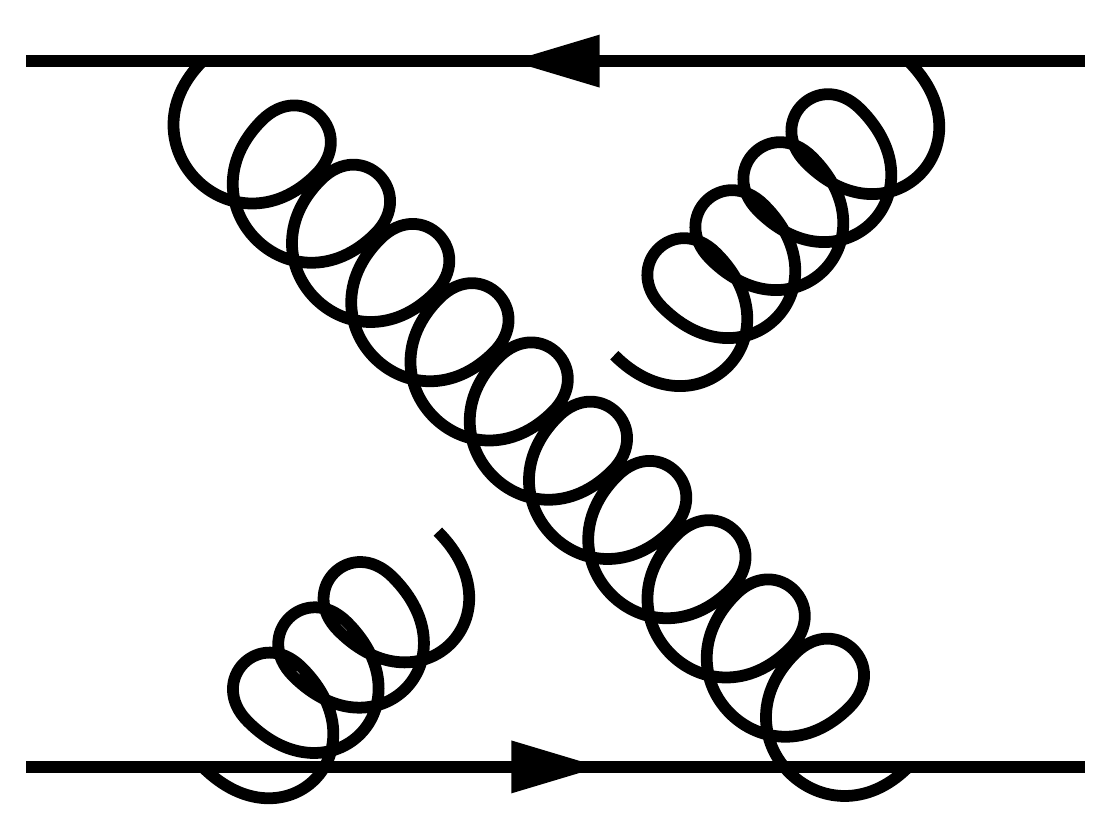}
    \caption{Example of diagrams that are not included in the ``ladder'' approximation scheme.}
    \label{fig:NonLadderApprox}
\end{figure}
In the ``ladder'' approximation scheme, we have dropped diagrams such as the crossed gluon lines shown in \Cref{fig:NonLadderApprox}.
Another type of approximation scheme beyond the ladder approximation is the ``non-ladder'' approximation.
This approximation scheme balances the color factors of some loops to make it resumable in a closed form, thus bringing in the milder quantitative systematic errors compared with the ``ladder'' case.
The mixed quark-gluon vertex then reads
\begin{align}
    (\mathcal A+\mathcal B)^{ab}_{\mu\nu} &\sim \bar{\psi} \biggl[ T^a T^b \Gamma_{\mu\nu}^A + T^b T^a \Gamma_{\mu\nu}^B \biggl]\psi \nn
    &= \bar{\psi} \biggl[ T^a T^b \Gamma_{\mu\nu}^A + T^a T^b \Gamma_{\mu\nu}^B \biggl]\psi \nn
    &\quad+ (\text{$[T^a,T^b]$ term}),
    \label{eq:nonLadderApproximationMainText}
\end{align}
where $\Gamma_{\mu\nu}^{A/B}$ denotes the tensorial structure apart from the color-space contribution.
We ignore the $[T^a, T^b]$ term in the beyond-ladder approximation scheme as an effect from the projection onto the fermionic subspaces by truncating the interaction term $\sim \partial_\mu F_{\mu\nu} \bar{\psi}\gamma_\nu\psi$ following the discussions in \cite{Aoki:2000dh,Aoki:2012mj}.

Then, after performing the Fierz transformation and several steps of algebraic calculations, the flow equation results in the form
\begin{align}
    &\partial_t V (\sigma;t) = \eta_\psi \sigma \partial_\sigma V(\sigma;t) \nn
    &\quad- \frac{1}{2} \left[ {\rm Flow}^{{\rm N.L.},-}_k(\sigma, M_\psi) + {\rm Flow}^{{\rm N.L.},+}_k(\sigma, M_\psi) \right].
    \label{eq:nonLadderFlowWithmuMaintext}
\end{align}
Because the explicit form of ${\rm Flow}^{{\rm N.L.},\pm}_k(\sigma, M_\psi)$ is lengthy, we do not exhibit it here. 
Instead, we give it in \cref{eq:defineFlowName} in \Cref{sec:LadderNonladder}.

\subsection{Gluonic sector and running gauge coupling}
\label{sec:GluonicSectorAndRunningCoupling}

\begin{figure}
    \centering
    \includegraphics[width=0.2\linewidth]{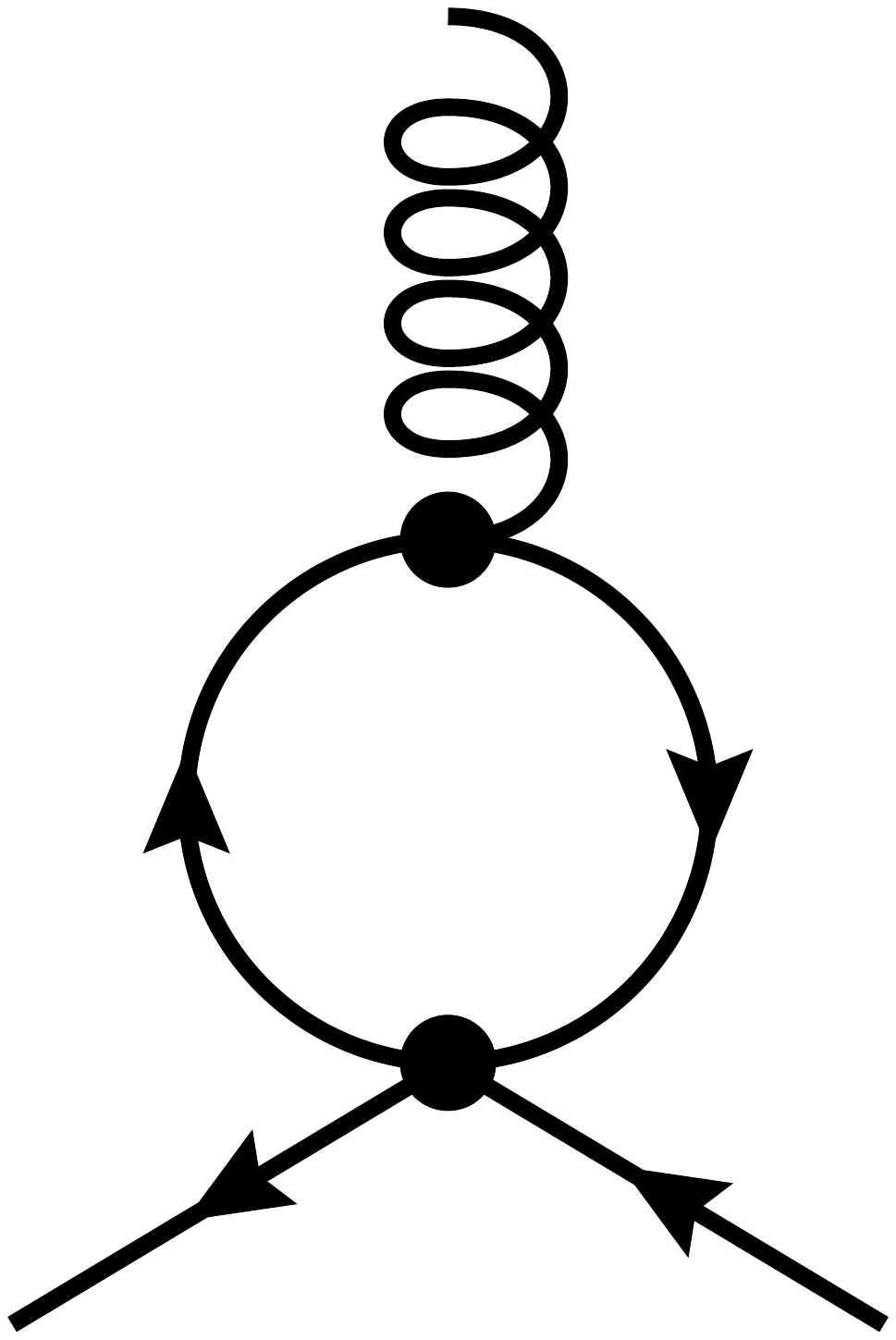}
    \caption{Feynman diagram which contributes to the running gauge coupling induced from the 4-fermion interaction.}
    \label{fig:partGaugeCoup}
\end{figure}

The flow equation of the fermionic potential $V(\sigma;t)$ is closed as a partial differential equation (PDE).
To evaluate it, we need to give the flow equation of the gauge coupling $g_s$.
%(also be regarded as the renormalized quark-gluon dressing in the first tensorial channel $\lambda^{(1)}$, see e.g., Ref.~\cite{Mitter:2014wpa}) and the gluon wave function renormalization $Z^A_k$.
As will be discussed in \Cref{sec:methodOfCharacteristics}, the flow equation of the fermionic potential, including the four-fermion coupling, is reformulated in a weak form to pass through the critical scale of dynamical chiral symmetry breaking, where the four-fermion coupling diverges. In contrast, the weak form of the flow equation for the gauge coupling, which receives contributions from the four-fermion coupling as shown in \Cref{fig:partGaugeCoup}, is not introduced. Consequently, the running gauge coupling encounters a divergence at the critical scale in this setup.
In earlier works~\cite{Higashijima:1983gx, Aoki:1990eq, Aoki:2012mj}, an IR cutoff was introduced for the running gauge coupling.
In the present study, instead of adopting this approach, we make use of quantitative results obtained from other functional methods~\cite{Gao:2024ggp, Gao:2020qsj, Lu:2023mkn, Lu:2025cls, Cyrol:2017ewj, Cyrol:2017qkl, Fu:2019hdw, Cyrol:2017ewj, Alkofer:2018guy}.
%%%%%%%%%%%%%%%%%%%%
%%%%%%%%%%%%%%%%%%%%

In this work, we take the minimal approach for the gluonic sectors, which also makes the ghost sector independent of the fermionic sector.
We take the $p = k$ prescription~\cite{Goertz:2024dnz} and hard thermal loop (HTL) approximation for the gluon wave function such that
\begin{align}
    Z^A_k = Z^{\rm trans}_{T,\mu_q}(p = k),
\end{align}
where the right-hand side denotes the transverse gluon propagator dressing at finite temperature and quark chemical potential, and 
\begin{align}
    Z^{{\rm trans}}_{T,\mu_q}(p) = Z_{\rm vac}(p) + \frac{4\pi}{3} \alpha_S^\text{HTL} \left( T^2 + \frac{3}{\pi^2} \mu_q^2 \right) \frac{1}{p^2},
    \label{eq:HTLmassFromQuarkLoop}
\end{align}
where $\alpha_S^\text{HTL} = 0.115$ is the gauge coupling for evaluating the HTL mass.
\Cref{eq:HTLmassFromQuarkLoop} implies that we only take into account the light quark vacuum polarization contribution to the thermal medium of the gluon, and the strange flavor contribution is dropped due to its large mass. 
For the vacuum gluon dressing $Z^{-1}_{\rm vac}(p)$, we take the ``functional-lattice'' propagator form \cite{Gao:2021wun,Lu:2023mkn}
\begin{align}
    &Z_{\rm vac}^{-1}(k^2) \nn
    &\quad = \frac{k^2\frac{a^2 + k^2}{b^2 + k^2}}{M_G^2(k^2) + k^2 \left[ 1 + c \log (d^2 k^2 + e^2 M_G^2(k^2)) \right]^\gamma},
    \label{eq:gluon propagator}
\end{align}
in which
\begin{align}
    M_G^2(k^2) = \frac{f^4}{g^2 + k^2},
    \label{eq:gluon mass}
\end{align}
and $\gamma = \frac{13-\frac{4}{3} N_f}{22 - \frac{4}{3}N_f}$ is the perturbative anomalous dimension of the gluon propagator.

Here we choose $N_f = 3$ to match the $(2+1)$-flavor result of the gluon propagator.
Within the current truncation scheme, the strange flavor and the light two-flavors are separated from each other due to the ignorance of the quantitative back-reaction in the gluon propagator.
Thus, this allows us to simulate the (2+1)-flavor result with only the light quark involved in evaluating the fermionic effective potential.

The values of the parameters are chosen as 
\begin{align}
&a= 1~{\rm GeV},&
&b=0.735~{\rm GeV},&
&c= 0.12,\nn
&d=0.0257~{\rm GeV}^{-1},&
&e=0.081~{\rm GeV}^{-1}.
\end{align}
in the gluon propagator \labelcref{eq:gluon propagator} and as
\begin{align}
&f=0.65~{\rm GeV},&
&g=0.87~{\rm GeV},
\end{align}
in the gluon mass \labelcref{eq:gluon mass}, to fit with the $(2+1)$-flavor data from the functional approaches~\cite{Gao:2021wun}.

For the gauge coupling before renormalization, we use the second-order Pad\'e approximation, which yields
\begin{align}
    \bar{g}_s(k) = \frac{0.2301 + 0.4411\,k + 0.3967 \,k^2}{0.0832 + 0.0838\,k + 0.4502\,k^2}.
\end{align}
to fit the data of the $\lambda^{(1)}_{\bar{q}Aq}(\bar{p})$ in the fRG approach to the (2+1) flavors \cite{Cyrol:2017ewj}.
Then, the renormalized gauge coupling reads
\begin{align}
    \alpha_s = \frac{1}{Z_k^A}\frac{\bar{g}_s^2}{4\pi}.
    \label{eq:renormalizedCoupling}
\end{align}

%%%%%%%%%%%%%%%%%%%%
\begin{figure}
    \centering
    \includegraphics[width=1\linewidth]{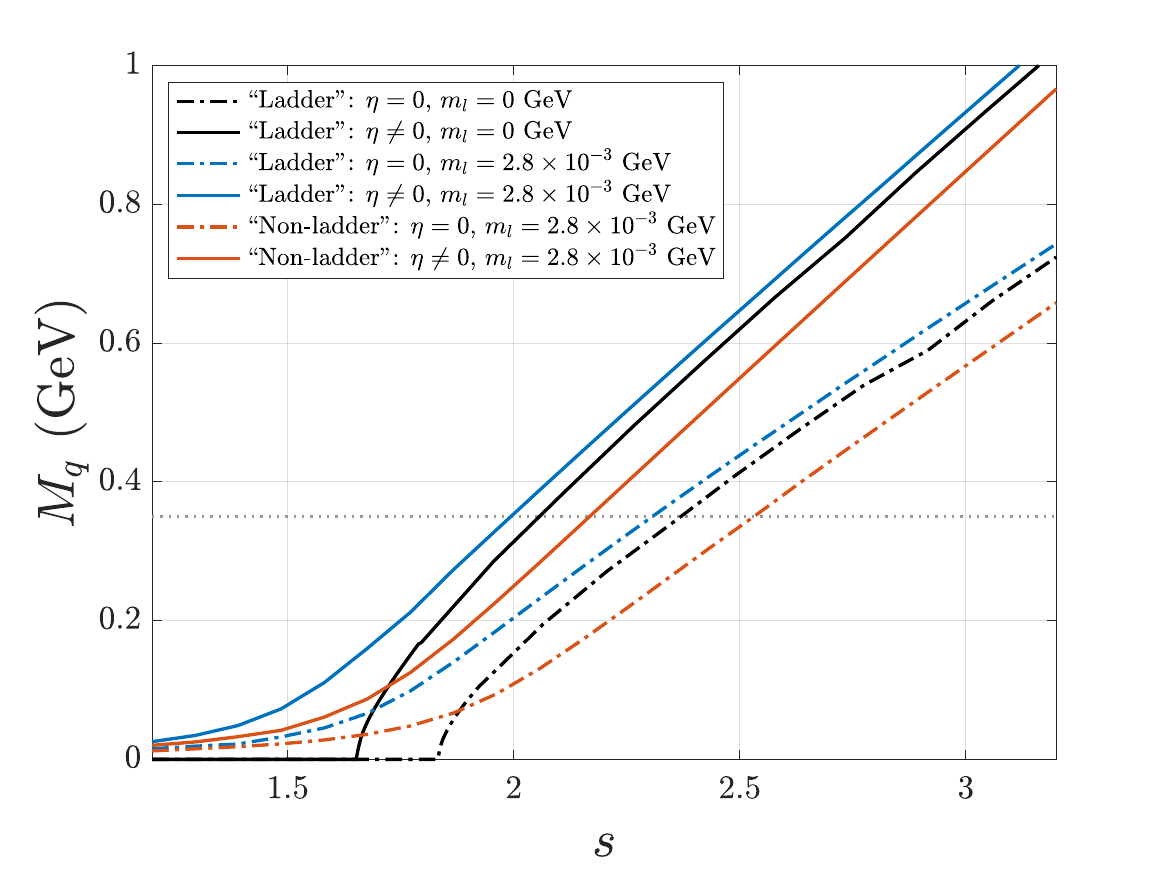}
    \caption{
    The scaling factor $s$-dependence of the constituent quark mass $M_q = \partial_\sigma V(\sigma = 0)$.
    }
    \label{fig:ConstMassScaling}
\end{figure}
%%%%%%%%%%%%%%%%%%%%

\begin{figure}
    \centering
    \includegraphics[width=1\linewidth]{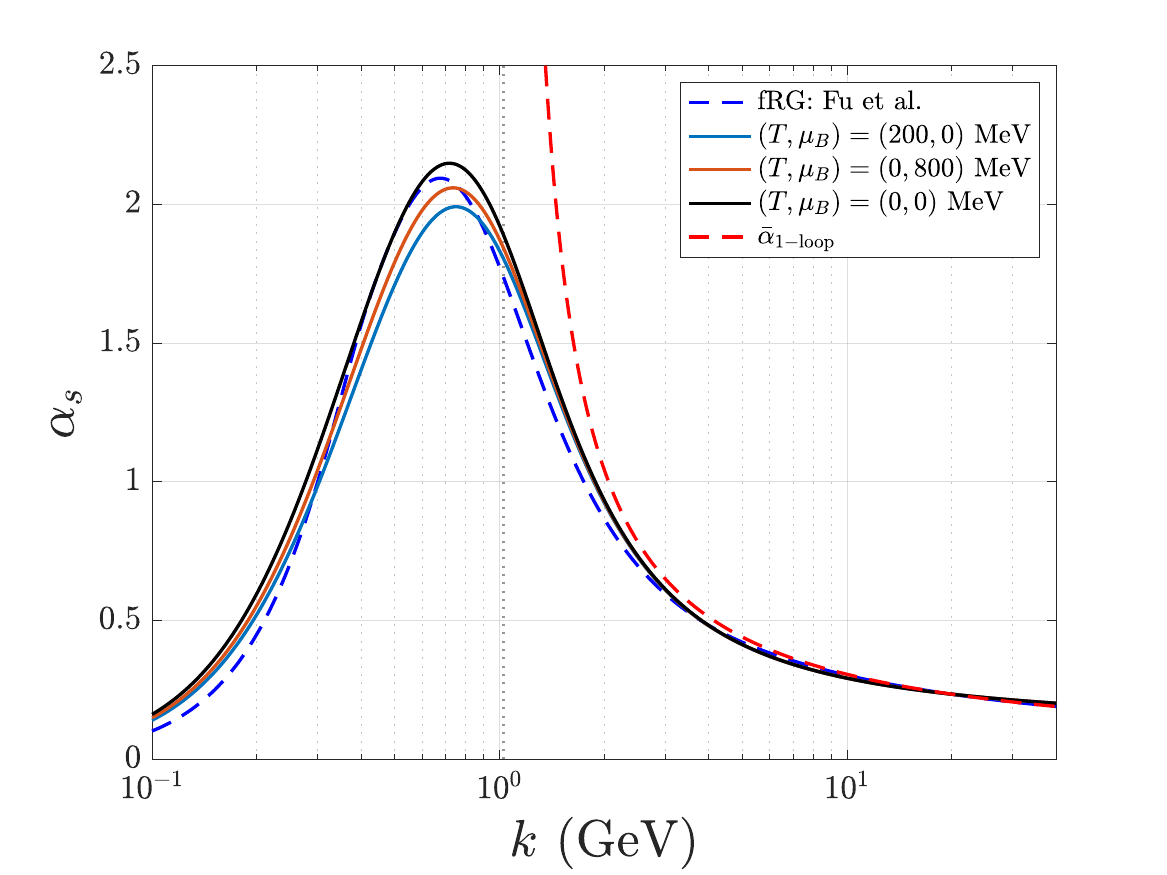}
    \caption{
    Rescaled running gauge coupling $\alpha_s$ (solid lines) defined in \cref{eq:renormalizedCoupling} as a function of the RG scale $k$, evaluated at different temperature $T$ and baryon chemical potential $\mu_B$, in comparison with the one obtained from the bosonized flow~\cite{Fu:2019hdw} (blue-dashed line), and the perturbative 1-loop running gauge coupling $\bar{\alpha}_\text{1-loop}$ (red-dashed line).
    The gray-dotted line denotes the position of the QCD scale $\Lambda_{\rm QCD}$ obtained from the IR Landau pole of $\bar{\alpha}_\text{1-loop}$.
    }
    \label{fig:gaugeCoupling}
\end{figure}

To obtain the phase diagram, the model parameters must be fixed so that reliable values of IR observables--such as the light constituent quark mass--are reproduced at zero temperature and zero chemical potential. In the present system, the gauge coupling serves as the central input parameter. However, a straightforward use of the input gauge coupling data does not reproduce the correct IR observables expected from lattice simulations. To control these IR observables, we introduce a rescaling factor that represents the degree of freedom associated with the gauge coupling\footnote{This treatment would be analogous to the introduction of an IR cutoff on the gauge coupling \cite{Higashijima:1983gx,Aoki:1990eq}. }
\begin{align}
    \alpha_s \rightarrow \bar{\alpha}_s = s \cdot \alpha_s.
\end{align}
For instance, $\bar{\alpha}_s(40~{\rm GeV}) = 0.1410$ for $s = 2.180$.
The dependence of the light constituent quark mass on the scaling factor is shown in \Cref{fig:ConstMassScaling} for several cases.
We plot the gauge coupling $\bar{\alpha}_s$ with the rescaling factor $s \simeq 3.13$ within the range of the RG scale $k \in [10^{-1},40]$~GeV in \Cref{fig:gaugeCoupling}, which corresponds to the renormalization condition $\alpha_s(20~{\rm GeV}) \simeq 0.235$ adopted in Ref.~\cite{Fu:2019hdw}.
In comparison to the perturbative one-loop running gauge coupling $\alpha_\text{1-loop}$, we choose the same renormalization condition as $\bar{\alpha}_\text{1-loop}(20~{\rm GeV}) = 0.235$, which is shown as the red solid line in \Cref{fig:gaugeCoupling}, and yields the QCD scale $\Lambda_{\rm QCD} \simeq 1.02~{\rm GeV}$.
This completes the current setup, and the flow equations of the fermionic potential are ready to be solved as closed PDEs.

%%%%%%%%%%%%%%%%%%%%
%%%%%%%%%%%%%%%%%%%%

%%%%%%%%%%%%%%%%%%%%%%%%%%%%%%%%%%%%%%%%%%%%%%%%%%%%%%%%%%%%%%%%%%%%%%
\section{Weak solution and method of characteristics}
\label{sec:WeakSolutionAndMethodOfCharacteristics}

We aim to obtain the dynamical mass (or chiral condensate) by solving the flow equations \labelcref{eq:ladderFlowMainText} and \labelcref{eq:nonLadderApproximationMainText}. 
However, the fermionic effective potential develops singularities (or discontinuities) at a certain energy scale as a result of the phase transition.
This implies that the flow of $V(\sigma; t)$ terminates at an intermediate scale, making it difficult to extract physical quantities in the deep infrared regime ($k \to 0$).
In this section, we introduce the strategy to solve the ladder~\labelcref{eq:ladderFlowMainText} and the non-ladder~\labelcref{eq:nonLadderApproximationMainText} flow equations with explicit discontinuity.
The method we will apply is within the framework of the weak solution \cite{Aoki:2014ola} in solving the first-order partial differential equations.

\subsection{Weak solution of the flow equation}

We start by considering the following first-order partial non-linear differential equation
\begin{align}\label{eq: strong PDE}
    \partial_t V(\sigma;t) = -F(\partial_\sigma V, \sigma ; t),
\end{align}
which could be the generalized version of the flow equations \labelcref{eq:ladderFlowMainText} and  \labelcref{eq:nonLadderApproximationMainText}.
When flowing the scale down into IR, the four-fermi coupling tends to be divergent since it corresponds to the chiral susceptibility for dynamical chiral symmetry breaking of second-order.
Here, defining the mass function and the four-fermi coupling with respect to the background field
\begin{align}
    M(\sigma;t) \equiv \partial_\sigma V (\sigma;t) , \quad G(\sigma;t) \equiv \partial_\sigma^2 V(\sigma;t),
\end{align}
the above fact can be translated into
\begin{align}
    |G(0^\pm; t_c^+)| \rightarrow \infty,
\end{align}
when the scale approaches the critical scale $t_c$.
Below the critical scale, the mass function (in the chiral limit) develops a discontinuity,
\begin{align}
     M(0^-;t) = - M(0^+;t) \neq 0,
     \label{eq:discontinuity}
\end{align}
which signals the onset of chiral symmetry breaking.

This discontinuity shows up during the integration of the RG flow and causes the flow to stop at that point.
Also, if the solution of the flow equation shows a discontinuity behavior \labelcref{eq:discontinuity}, the solution cannot be described by the original PDE.
In the case of a second-order phase transition, such a discontinuity appears at $\sigma = 0$, whereas a first-order phase transition, which may occur at finite chemical potential, involves discontinuities at nonzero values of $\sigma$.
Thus, the PDE must be reformulated so as to accommodate solutions with singularities.

Let us now introduce the weak formulation for the PDE.
To this end, we consider the flow equation for $M(\sigma; t)$ instead of that for $V(\sigma; t)$.
Taking the derivative with respect to $\sigma$ on both sides of \cref{eq: strong PDE}, we obtain the PDE of the mass function
\begin{align}
    \partial_t M(\sigma; t) &= -\partial_\sigma F\left( M(\sigma; t), \sigma ; t \right) \nn
    &= -\frac{\partial F}{\partial M} \cdot \frac{\partial M}{\partial \sigma} - \frac{\partial F}{\partial \sigma}.
\end{align}
This equation is, in general, a conservation law, where the conserved ``charge'' is $M$ and the associated ``current'' is $F$. A typical example of such a conservation equation is the Burgers equation in hydrodynamics~\cite{Burgers1995}.
We integrate the PDE of the mass function multiplied by an arbitrary smooth function $\varphi(\sigma;t)$, which satisfies
\begin{align}
    \lim_{\sigma\rightarrow \pm \infty} \varphi(\sigma;t) = 0 , \quad \lim_{t \rightarrow -\infty} \varphi(\sigma;t) = 0,
\end{align}
and utilizing the integration by parts, we obtain the following form
\begin{align}\label{eq: weak PDE}
    \int_{-\infty}^0 \df t \int_{-\infty}^\infty \df \sigma \, \left( M \frac{\partial \varphi}{\partial t} + F \frac{\partial \varphi}{\partial \sigma} \right) = - \int_{-\infty}^\infty \df \sigma \, \left( M \, \varphi \right)_{t=0}.
\end{align}
Since the derivatives with respect to $\sigma$ and $t$ are transferred onto the test function $\varphi$, the solution $M$ is not required to be differentiable. We refer to \cref{eq: weak PDE} as the \textit{weak flow equation}, and its solution is called the \textit{weak solution}. For details on the weak solution, see Refs.~\cite{Aoki:2014ola,evans2010partial}.

%%%%%%%%%%%%%%%%%%%%%%%%%%%%%%%%%%%%%%%%%%%%%%%%%%%
\begin{figure*}
    \centering
    \includegraphics[width=0.47\linewidth]{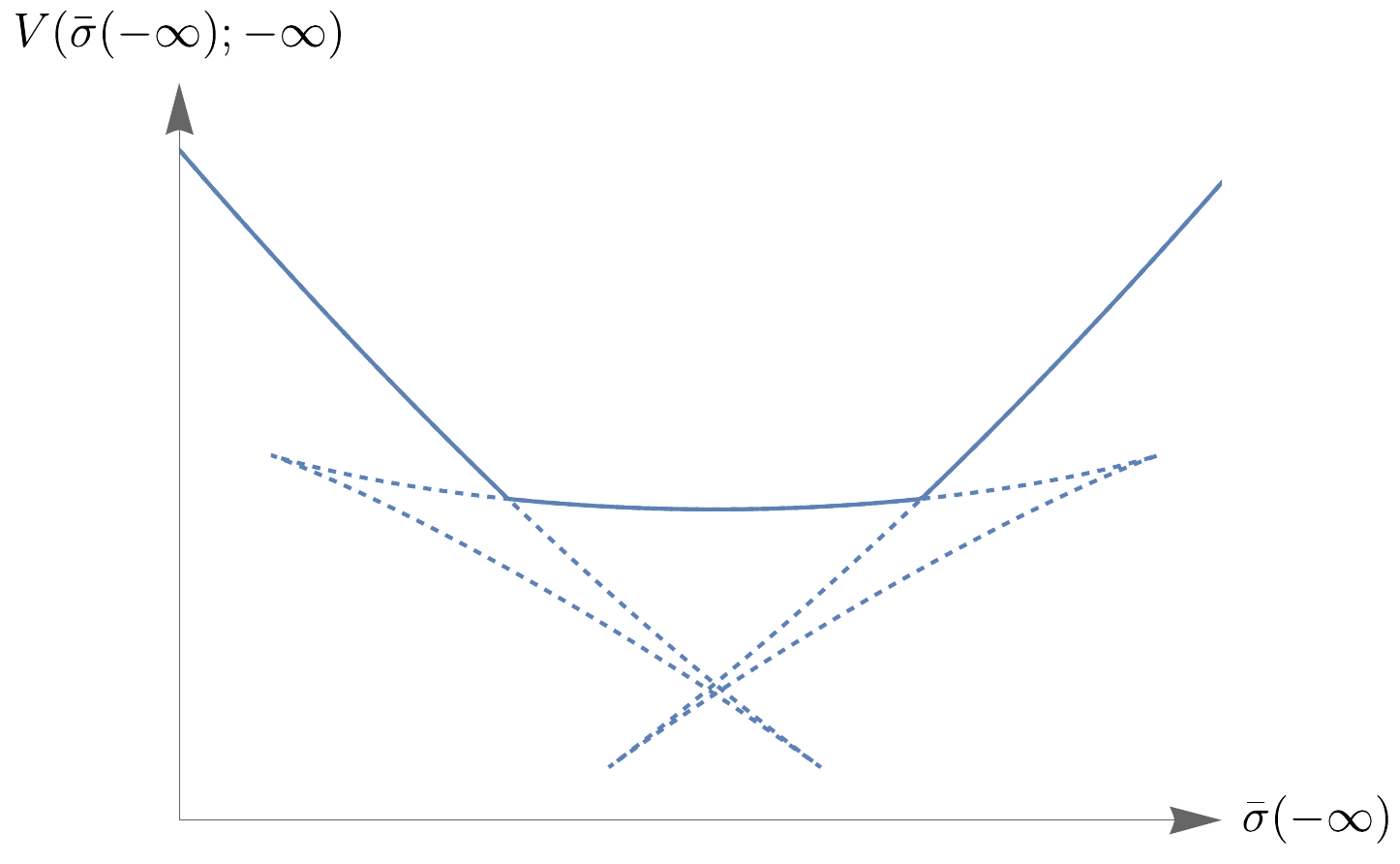}
    \quad
    \includegraphics[width=0.47\linewidth]{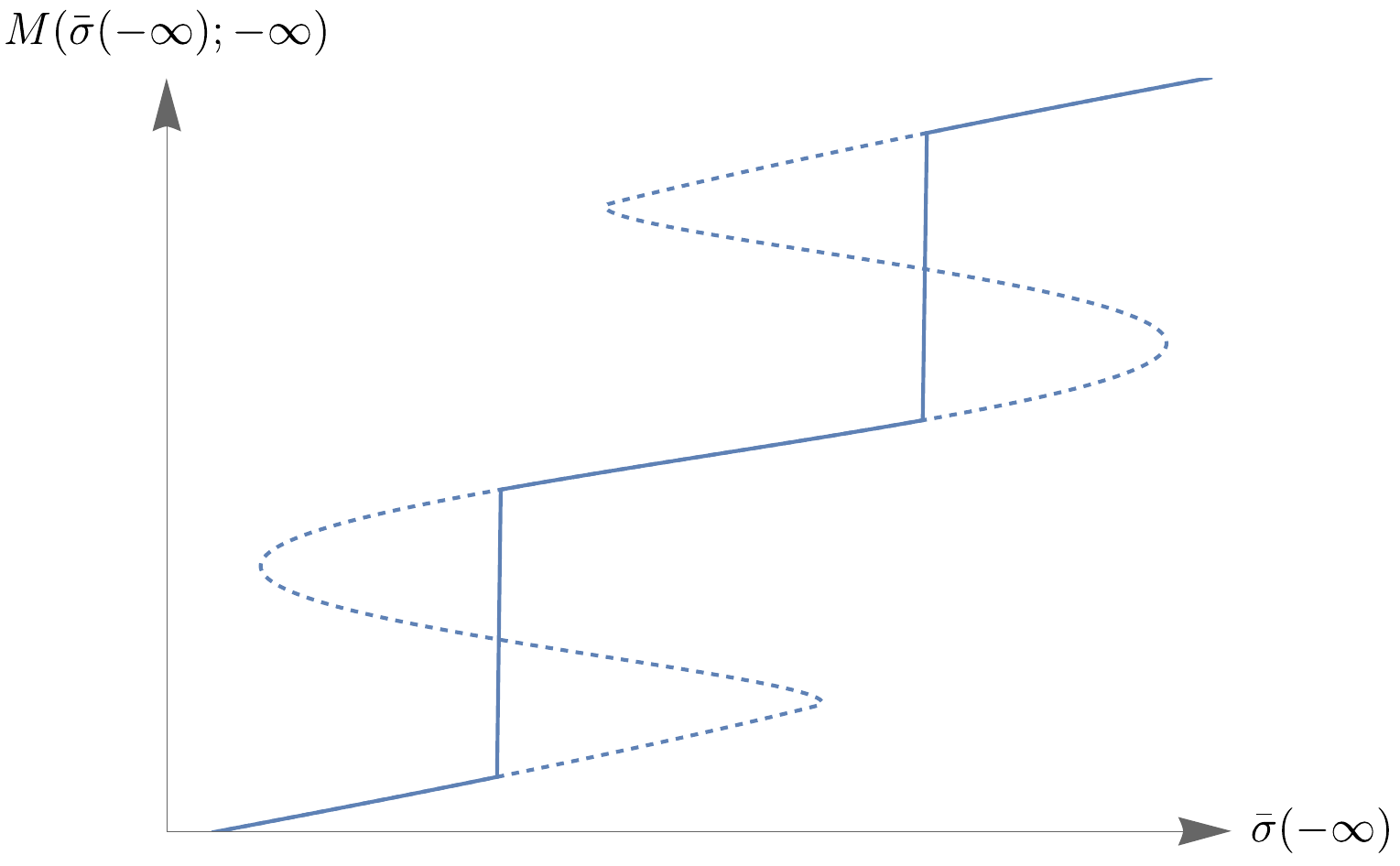}
    \caption{
    Schematic sketch of the RH condition \labelcref{eq: Rankine–Hugoniot condition} in searching for the weak solution.
    The right panel shows the fermionic mass function $M(\bar{\sigma};t)$ solved from the method of characteristics, which will be explained in \Cref{sec:methodOfCharacteristics} and shows a multivalued behavior.
    The left panel is the corresponding fermionic potential $V(\bar{\sigma};t)$.
    Following the discussion in \Cref{sec:RHCondition}, we drop the dashed part in $V(\bar{\sigma};t)$, resulting in the solid line in both panels, shows the non-analyticity of $V(\bar{\sigma};t)$ and discontinuity of $M(\bar{\sigma};t)$.
    }
    \label{fig:WeakExampleNJL}
\end{figure*}
%%%%%%%%%%%%%%%%%%%%%%%%%%%%%%%%%%%%%%%%%%%%%%%%%%%

\subsection{Rankine-Hugoniot condition}
\label{sec:RHCondition}

Next, we discuss the requirement of the singularity of the weak solution $M = M(\sigma;t)$ following the discussion of Ref.~\cite{Aoki:2014ola}.
Suppose the set of discontinuous points of the weak solution forms a curve $\sigma^*(t)$ in the $(\sigma,t)$ plane, which divides the weak solution into two regions on different sides, denoted as $M^{(L)}$ and $M^{(R)}$.
According to the conservation law nature of the original equation, if we take a small area made of the variation of $\sigma^*$ and $t$, at the boundary of this area we have
\begin{align}\label{eq: Rankine–Hugoniot condition}
    \left(M^{(L)} - M^{(R)}\right) \df \sigma^* = \left[ F(M^{(L)}) - F(M^{(R)}) \right] \df t.
\end{align}
This condition is the Rankine–Hugoniot (RH) condition of the weak solution $M(\sigma;t)$.
For the convenience of finding the weak solution, we shall consider one step further.
We denote the limit values of the fermionic potential at the different side of the singularity trajectory $\sigma^*$ as $V^{(L)}$ and $V^{(R)}$, and we trace the variation of them $\df V^{(L/R)}(\sigma^*(t))$ on the trajectory, which reads from the original RG equation
\begin{align}
    \df V^{(L/R)}(\sigma^*(t)) &= \left. \frac{\partial V}{\partial \sigma} \right|_{(L/R)} \df \sigma^* + \left. \frac{\partial V}{\partial t} \right|_{(L/R)} \df t \nonumber\\
    &= M^{(L/R)} \df \sigma^* - F(M^{(L/R)}) \df t.
\end{align}
Then, the difference between $\df V^{(L/R)}(\sigma^*(t);t)$ vanishes according to the RH condition~\labelcref{eq: Rankine–Hugoniot condition}
\begin{align}
    \df V^{(L)}(\sigma^*(t);t) - \df V^{(R)}(\sigma^*(t);t) = 0.
\end{align}
Since the initial potential continues everywhere within our setup, this continuity will be kept even after the emergence of the singularity trajectory.
Thus we have 
\begin{align}
    V^{(L)}(\sigma^*(t);t) = V^{(R)}(\sigma^*(t);t).
\end{align}

Here, we show a schematic example of the weak solution to the weak flow equation in the Nambu--Jona-Lasinio model at finite density and temperature~\cite{Aoki:2017rjl} in \Cref{fig:WeakExampleNJL}.
On the left panel of \Cref{fig:WeakExampleNJL}, we show the fermionic potential of the composite operator $V(\sigma;t)$ at deep IR, which behaves with two singularity points.
Imposing the continuity condition of the fermionic potential, we take the superior values of this potential, and the weak solution would be obtained as in the right panel.
The method of finding this solution will be introduced in the next subsection.

\subsection{Method of characteristics}
\label{sec:methodOfCharacteristics}

Following the method of characteristics, we introduce the characteristic curves $\bar{\sigma}(t)$ with initial conditions $\bar{\sigma}(t=0) = \sigma_0$, which satisfy
\begin{align}\label{eq: dsigmadt}
    \frac{\df}{\df t} \bar{\sigma}(t) = \frac{\partial F(\bar{M}, \bar{\sigma};t)}{\partial \bar{M}},
\end{align}
where $\bar{M} = M(\bar{\sigma}(t);t)$.
With the introduction of $\bar{\sigma}(t)$, we convert the original RG equation into an equivalent strong form by looking at
\begin{align}\label{eq: dMdt}
    \frac{\df}{\df t} \bar{M} &= \left. \frac{\partial M(\sigma;t)}{\partial \sigma} \right|_{\sigma = \bar{\sigma}} \frac{\df}{\df t} \bar{\sigma} + \frac{\partial M(\bar{\sigma};t)}{\partial t} \nn
    &= -\left. \frac{\partial F(M, \sigma;t)}{\partial \sigma} \right|_{\sigma = \bar{\sigma}}.
\end{align}
The coupled ordinary differential equations (ODEs) \labelcref{eq: dsigmadt} and \labelcref{eq: dMdt} give a set of solutions $(\bar{\sigma}(t), M(\bar{\sigma}(t);t))$ at different scale $t$ with different initial conditions $\bar{\sigma}(t=0) = \sigma_0$ and $M(\bar{\sigma}(0);0)) = M(\sigma_0;0))$, and they forms a unique strong solution of the RG equation.
The fermionic potential along the characteristic curve is integrated from the original RG equation
\begin{align}
    V(\bar{\sigma}(t);t) &= V(\sigma_0;0) + \int_0^t \df t^\prime \left[ \frac{\partial V}{\partial \bar{\sigma}} \frac{\df \bar{\sigma}}{\df t^\prime} + \frac{\partial V}{\partial t} \right] \nonumber\\
    &= V(\sigma_0;0) + \int_0^t \df t^\prime \left[ \bar{M} \frac{\df \bar{\sigma}}{\df t^\prime} - F(\bar{M}, \bar{\sigma};t^\prime) \right].
    \label{eq:equationOfV}
\end{align}

Following the discussion on the previous example of the RH condition in \Cref{fig:WeakExampleNJL}, we select the initial conditions $\{\sigma_0\}$ such that the fermionic potential takes the superior values in the vicinity of each point $\{\bar{\sigma}\}$.
Then we trace the mass function $\bar{M}$ from the selected initial points, and we obtain the weak solution like in the right panel of \Cref{fig:WeakExampleNJL}.
This completes the current discussion on the search for the weak RG solution, which preserves the divergence nature of the four-fermi coupling in the chiral broken phase.

%%%%%%%%%%%%%%%%%%%%%%%%%%%%%%%%%%%%%%%%%%%%%%%%%%%%%%%%%%%%%%%%%%%%%%
\section{Results}
\label{sec:results}

In this section, we show our numerical results from solving the flow equations of the fermionic potential \labelcref{eq:ladderFlowMainText} and \labelcref{eq:nonLadderApproximationMainText} utilizing the methods described in \Cref{sec:WeakSolutionAndMethodOfCharacteristics} and \Cref{sec:numericalDetails}.
With the parameters fixed by physical quantities, we investigate the phase diagram through the dynamical mass $\bar{M}(0;-\infty)$.

\subsection{Numerical implementation}
\label{sec:numericalDetails}

Before presenting the numerical results, we briefly outline the numerical setup for completeness.
At the initial RG scale $\Lambda$, we solve the coupled ODEs \labelcref{eq: dsigmadt} and \labelcref{eq: dMdt} using the $5$-th order Runge-Kutta  method with adaptive step size, with the initial condition
\begin{align}
    \bar{\sigma}(0) = \sigma_0, \quad \bar{M}(\bar{\sigma}(0);0) = m_l.
\end{align}
The Matsubara frequencies in the flow equation are summed explicitly up to a large momentum scale, e.g., the initial scale $\Lambda$.
At each Runge-Kutta step, we also couple the flow equation with \cref{eq:equationOfV} to solve the fermionic potential.
We search for a wide range of the initial value of $\sigma_0$ to fully cover the multi-valued region of $\bar{M}$ at each scale $t \in [-11,0]$, and extracting the value of $\bar{M}(0,t)$ by the linear interpolation of the nearest points around $\bar{\sigma} = 0$ from the RH condition~\labelcref{eq: Rankine–Hugoniot condition}.

\subsection{Quark mass function and parameter fixing}

We first describe how the parameters are fixed within the current setup.
In this work, we have two free parameters: the enhancement factor $s$ reflecting the gauge coupling at the UV scale according to the discussion around \cref{eq:renormalizedCoupling}, and the light current quark mass $m_l$.
The intrinsic scale $\Lambda_\text{QCD}$ is already encoded in the input of the gluon propagator, see the discussion in \Cref{sec:GluonicSectorAndRunningCoupling}.
The computation is done in the Landau gauge $\xi = 0$ and $k\in (0,40]~{\rm GeV}$ to match the RG conditions of the input data.

We fix those two parameters by fitting the physical quantities of the $\sigma^0$-meson mass $m_\sigma \sim 700$~MeV and the $\pi^0$-meson mass $m_\pi \sim 135$~MeV at vanishing temperature and quark chemical potential, through the following relations
\begin{align}
    m_\sigma \sim 2 \bar{M}(0;-\infty) \equiv 2M_q, \quad m_\pi^2 f_\pi^2 \sim 2 m_l \langle \bar{\psi}\psi \rangle,
\end{align}
where $f_\pi$ is the pion decay constant and $\langle \bar{\psi}\psi \rangle$ is the light quark condensate.

\begin{figure}
    \centering
    \includegraphics[width=0.9\linewidth]{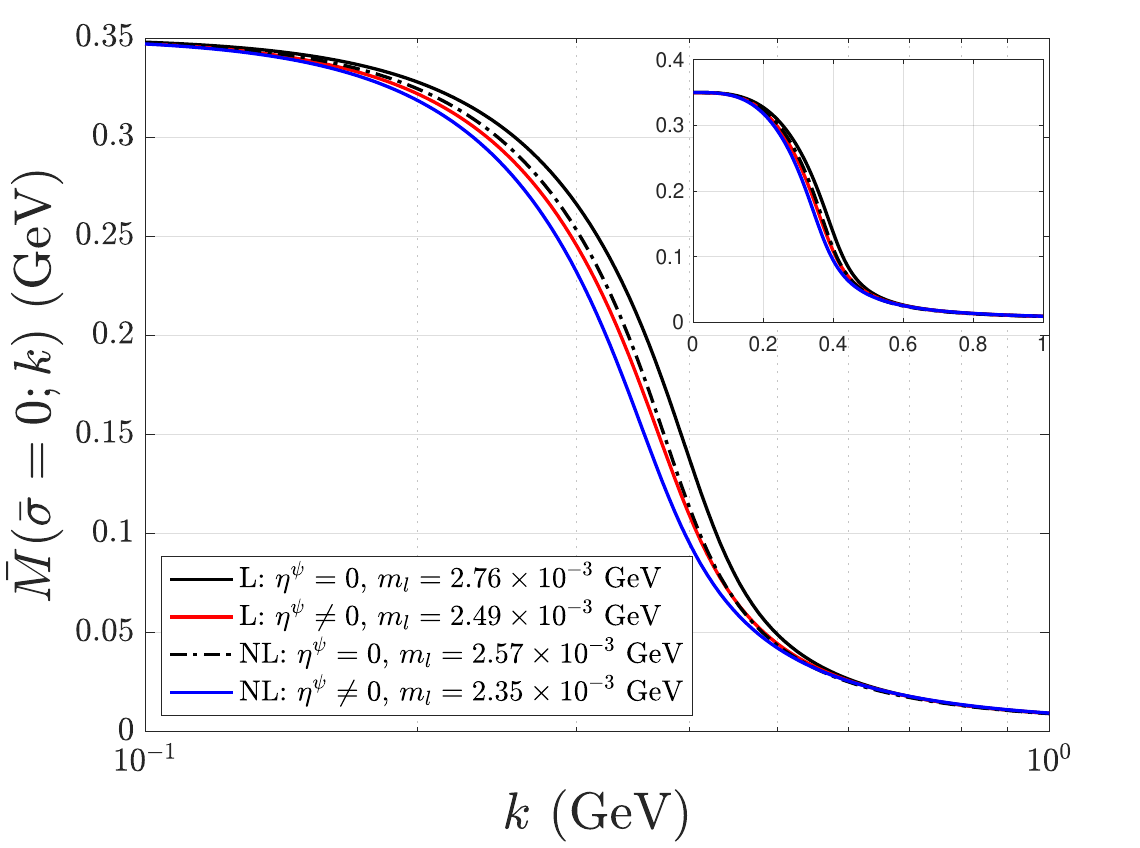}
    \caption{
    Quark mass function $\bar{M}(0;t)$ with respect to the RG scale $k=\Lambda e^{t}$ at the vacuum.
    We identify this mass function with the momentum argument as stated in \cref{eq:massFunctionReadOff}.
    }
    \label{fig:massFunctions}
\end{figure}

To compute the value of $f_\pi$ and $\langle \bar{\psi}\psi \rangle$, we read off the quark mass function through the following identification
\begin{align}
    M_q(p) = \bar{M}(0;k=p),
    \label{eq:massFunctionReadOff}
\end{align}
i.e., the replacement of the momentum argument with the RG scale $k$.
The corresponding mass function $M_q(p)$ for $p \in [k_\text{IR},\Lambda] = [10^{-3},40]~{\rm GeV}$ is shown in \Cref{fig:massFunctions} for the ``ladder'' and ``non-ladder'' type of flow equations with/without the quark anomalous dimension at the physical point.

Then, the pion decay constant $f_\pi$ is obtained through the reduced Pagels-Stokar formula
\begin{align}
    f_\pi^2 = \frac{3}{\pi^2}\int_{k_\text{IR}}^\Lambda \df p \frac{p^3 M_q(p)}{\left[ p^2 + M_q^2(p) \right]^2} \left[ M_q(p) - \frac{p}{4} \partial_p M_q(p) \right],
    \label{eq:Pagels-Stokar formula}
\end{align}
and the light quark condensate is given by
\begin{align}
    \langle \bar{\psi}\psi \rangle = -\frac{N_c}{2\pi^2} \int_{k_\text{IR}}^\Lambda \df p\, p^3 \left[ \frac{M_q(p)}{p^2+ M_q^2(p)} - \frac{m_l}{p^2+m_l^2}  \right].
\end{align}
Then, we fix the constituent quark mass $M_q = M_q(0) \sim 350~{\rm MeV}$ and scale the current quark mass, then the parameters are chosen as in \Cref{tab:parameterFixing}.
We note that using the Pagels–Stokar formula \labelcref{eq:Pagels-Stokar formula} introduces a sizable systematic error in evaluating the pion mass.
For example, in the case of ``non-ladder'' with quark anomalous dimension, the value of $f_\pi$ deviates from the experimental value $f_\pi^{\rm ex} \sim 92.1$ by about 15.2\%.
This systematic error also arises from the truncation scheme we adopt in this work; thus, we show our result as both a benchmark and a demonstration.
\begin{table}[]
\begin{tabular}{cc|c|c|c|c}
\hline \hline
\multicolumn{2}{c|}{Case}                                                                                                             & Parameters       & Values & Observables & \begin{tabular}[c]{@{}c@{}}Values\\ (MeV)\end{tabular} \\ \hline\hline
\multicolumn{1}{c|}{\multirow{10}{*}{``Ladder''}}                                               & \multirow{5}{*}{$\eta_\psi = 0$}    & $m_l$ (MeV)      & 0      & $M_q$       & 350.4                                                  \\ \cline{3-6} 
\multicolumn{1}{c|}{}                                                                           &                                     & $\bar{\alpha}_s$ & 0.1529 &             &                                                        \\ \cline{3-6} 
\multicolumn{1}{c|}{}                                                                           &                                     & $m_l$ (MeV)      & 2.76   & $M_q$       & 350.2                                                  \\ \cline{3-6}
\multicolumn{1}{c|}{}                                                                           &                                     & $\bar{\alpha}_s$ & 0.1493 & $m_\pi$     & 134.9                                                  \\ \cline{3-6} 
\multicolumn{1}{c|}{}                                                                           &                                     &                  &        & $f_\pi$     & 83.2                                                   \\ \cline{2-6} 
\multicolumn{1}{c|}{}                                                                           & \multirow{5}{*}{$\eta_\psi \neq 0$} & $m_l$ (MeV)      & 0      & $M_q$       & 351.1                                                  \\ \cline{3-6} 
\multicolumn{1}{c|}{}                                                                           &                                     & $\bar{\alpha}_s$ & 0.1331 &             &                                                        \\ \cline{3-6} 
\multicolumn{1}{c|}{}                                                                           &                                     & $m_l$ (MeV)      & 2.49   & $M_q$       & 350.1                                                  \\ \cline{3-6} 
\multicolumn{1}{c|}{}                                                                           &                                     & $\bar{\alpha}_s$ & 0.1294 & $m_\pi$     & 135.3                                                  \\ \cline{3-6} 
\multicolumn{1}{c|}{}                                                                           &                                     &                  &        & $f_\pi$     & 80.0                                                   \\ \hline
\multicolumn{1}{c|}{\multirow{6}{*}{\begin{tabular}[c]{@{}c@{}}``Non-\\ ladder''\end{tabular}}} & \multirow{3}{*}{$\eta_\psi = 0$}    & $m_l$ (MeV)      & 2.57   & $M_q$       & 350.3                                                  \\ \cline{3-6} 
\multicolumn{1}{c|}{}                                                                           &                                     & $\bar{\alpha}_s$ & 0.1640 & $m_\pi$     & 135.0                                                  \\ \cline{3-6} 
\multicolumn{1}{c|}{}                                                                           &                                     &                  &        & $f_\pi$     & 80.6                                                   \\ \cline{2-6} 
\multicolumn{1}{c|}{}                                                                           & \multirow{3}{*}{$\eta_\psi \neq 0$} & $m_l$ (MeV)      & 2.35   & $M_q$       & 350.3                                                  \\ \cline{3-6} 
\multicolumn{1}{c|}{}                                                                           &                                     & $\bar{\alpha}_s$ & 0.1410 & $m_\pi$     & 136.2                                                  \\ \cline{3-6} 
\multicolumn{1}{c|}{}                                                                           &                                     &                  &        & $f_\pi$     & 78.1                                                   \\ \hline \hline
\end{tabular}
\caption{Parameter fixing in this work. We present results for both the ``ladder" and ``non-ladder" cases, with and without the quark anomalous dimension, and also the case of the chiral limit in the ``ladder'' case.}
\label{tab:parameterFixing}
\end{table}

\subsection{Phase diagrams}

In \Cref{fig:TogetherPhaseDiagram}, we show our result of the QCD phase diagram in the $(\mu_B,T)$ plane, with the different parameter choices listed in \Cref{tab:parameterFixing}.
In the chiral limit ($m_l = 0$), the phase boundary is observed at $M_q = 0$, reflecting the second-order phase transition nature in the small $\mu_B$ region.
At the physical point, the phase transition then becomes the crossover.
We characterize the (pseudo) critical temperature line ($T_c$-line) by the maximum of the quantity $-\partial_T M_q$ at fixed baryon chemical potential, which is a similar quantity to the thermal susceptibility defined by the quark condensate $\chi_T = -\partial_T \langle \bar\psi\psi\rangle$, since the latter encountered with complicated subtraction schemes and special numerical treatment, which is beyond the scope of this work.
We then summarize the critical temperature at zero baryon chemical potential $T_c(0)$, the curvature $\kappa$ of the $T_c$-line around $\mu_B = 0$, and the critical end point (CEP) $(T_c,\mu_{B,c})$ in \Cref{tab:phaseDiagrams}.
Here, the curvature $\kappa$ is defined by the following expansion
\begin{align}
    \frac{T_c(\mu_B)}{T_c(0)} = 1 - \kappa \left( \frac{\mu_B}{T_c(0)} \right)^2 + \cdots.
\end{align}

Compared with the phase diagrams obtained from the bosonized fRG approach~\cite{Fu:2019hdw} and the DSE approaches~\cite{Gao:2020qsj, Fischer:2014ata}, the results in this work show the following characteristic behaviors:
\begin{itemize}
    \item The position of the CEP is far in the chiral limit of the ``ladder'' case, which becomes more severe when the current quark mass comes in.
    Also, we cannot even observe the signal of the second- or first-order phase transition in the cases at the physical point without the quark anomalous dimension down to the temperature range around $10$~MeV.

    \item The value of $T_c(0)$ in the “ladder” cases is slightly higher than that obtained from other functional methods for the $(2+1)$-flavor system, which typically yield values around 155~MeV~\cite{Gao:2020qsj}. In contrast, the “non-ladder” case with the inclusion of the quark anomalous dimension gives $T_c(0) = 158.0$~MeV, which is relatively close to the 155~MeV benchmark.

    \item The curvature $\kappa$ of the ``ladder'' cases is high in comparison with, e.g., $\kappa = 0.0150(7)$~\cite{Gao:2020qsj}.
    We infer that this difference originates from the quark vacuum polarization contribution to the gluon propagator, indicating that the HTL mass $\sim \mu_q^2$ is too large compared to the quantitative approaches.
\end{itemize}
From the above observations, we conclude that mesonic fluctuations play a subleading role compared to quantitative corrections to the gluonic loop, including the ``non-ladder'' types of diagrams around $\mu_B = 0$.
Meanwhile, the mesonic fluctuations become necessary in realizing the chiral criticality with moderate baryon chemical potential.

\begin{table}[]
\begin{tabular}{ccc|c|c}
\hline \hline
\multicolumn{3}{c|}{Cases}                                                                                                                       & Observables                                                       & Values        \\ \hline\hline
\multicolumn{1}{c|}{\multirow{12}{*}{``Ladder''}}    & \multicolumn{1}{c|}{\multirow{6}{*}{$\eta_\psi = 0$}}    & \multirow{3}{*}{$m_l = 0$}    & $T_c(0)$ (MeV)                                                    & 178.8         \\ \cline{4-5} 
\multicolumn{1}{c|}{}                                & \multicolumn{1}{c|}{}                                    &                               & $\kappa$                                                          & 0.0197        \\ \cline{4-5} 
\multicolumn{1}{c|}{}                                & \multicolumn{1}{c|}{}                                    &                               & \begin{tabular}[c]{@{}c@{}}$(T_c,\mu_{B,c})$\\ (MeV)\end{tabular} & (94.16,874.6) \\ \cline{3-5} 
\multicolumn{1}{c|}{}                                & \multicolumn{1}{c|}{}                                    & \multirow{3}{*}{$m_l \neq 0$} & $T_c(0)$ (MeV)                                                    & 176.8         \\ \cline{4-5} 
\multicolumn{1}{c|}{}                                & \multicolumn{1}{c|}{}                                    &                               & $\kappa$                                                          & 0.0193        \\ \cline{4-5} 
\multicolumn{1}{c|}{}                                & \multicolumn{1}{c|}{}                                    &                               & \begin{tabular}[c]{@{}c@{}}$(T_c,\mu_{B,c})$\\ (MeV)\end{tabular} & -  \\ \cline{2-5} 
\multicolumn{1}{c|}{}                                & \multicolumn{1}{c|}{\multirow{6}{*}{$\eta_\psi \neq 0$}} & \multirow{3}{*}{$m_l = 0$}    & $T_c(0)$ (MeV)                                                    & 169.6         \\ \cline{4-5} 
\multicolumn{1}{c|}{}                                & \multicolumn{1}{c|}{}                                    &                               & $\kappa$                                                          & 0.0183        \\ \cline{4-5} 
\multicolumn{1}{c|}{}                                & \multicolumn{1}{c|}{}                                    &                               & \begin{tabular}[c]{@{}c@{}}$(T_c,\mu_{B,c})$\\ (MeV)\end{tabular} & (93.31,813.6) \\ \cline{3-5} 
\multicolumn{1}{c|}{}                                & \multicolumn{1}{c|}{}                                    & \multirow{3}{*}{$m_l \neq 0$} & $T_c(0)$ (MeV)                                                    & 166.3         \\ \cline{4-5} 
\multicolumn{1}{c|}{}                                & \multicolumn{1}{c|}{}                                    &                               & $\kappa$                                                          & 0.0187        \\ \cline{4-5} 
\multicolumn{1}{c|}{}                                & \multicolumn{1}{c|}{}                                    &                               & \begin{tabular}[c]{@{}c@{}}$(T_c,\mu_{B,c})$\\ (MeV)\end{tabular} & (47.22,1058)  \\ \hline
\multicolumn{1}{c|}{\multirow{6}{*}{``Non-ladder''}} & \multicolumn{1}{c|}{\multirow{3}{*}{$\eta_\psi = 0$}}    & \multirow{6}{*}{$m_l \neq 0$} & $T_c(0)$ (MeV)                                                    & 166.1         \\ \cline{4-5} 
\multicolumn{1}{c|}{}                                & \multicolumn{1}{c|}{}                                    &                               & $\kappa$                                                          & 0.0176        \\ \cline{4-5} 
\multicolumn{1}{c|}{}                                & \multicolumn{1}{c|}{}                                    &                               & \begin{tabular}[c]{@{}c@{}}$(T_c,\mu_{B,c})$\\ (MeV)\end{tabular} & -  \\ \cline{2-2} \cline{4-5} 
\multicolumn{1}{c|}{}                                & \multicolumn{1}{c|}{\multirow{3}{*}{$\eta_\psi \neq 0$}} &                               & $T_c(0)$ (MeV)                                                    & 158.0         \\ \cline{4-5} 
\multicolumn{1}{c|}{}                                & \multicolumn{1}{c|}{}                                    &                               & $\kappa$                                                          & 0.0198        \\ \cline{4-5} 
\multicolumn{1}{c|}{}                                & \multicolumn{1}{c|}{}                                    &                               & \begin{tabular}[c]{@{}c@{}}$(T_c,\mu_{B,c})$\\ (MeV)\end{tabular} & (35.85,1078)  \\ \hline \hline
\end{tabular}
\caption{Observables related to the QCD phase diagram in \Cref{fig:TogetherPhaseDiagram}.}
\label{tab:phaseDiagrams}
\end{table}
%

%%%%%%%%%%%%%%%%%%%%%%%%%%%%%%%%%%%%%%%%%%%%%%%%%%%%%%%%%%%%%%%%%%%%%%
\section{Summary and conclusion}
\label{sec:summary}

In this paper, we have studied the QCD phase diagram without explicitly including mesonic fluctuations, employing the weak solution of the functional renormalization group (fRG) equation.  
The aim of this work is to revisit the well-studied problem of the QCD phase diagram with light quark flavors and to elucidate the role of mesonic fluctuations by focusing solely on quark and gluon dynamics.  
In addition, we investigated the impact of the ``non-ladder'' approximation scheme on the QCD phase transition in the vicinity of vanishing baryon chemical potential.

As an approximation, we introduced the ansatz \( V_k(\psi, \bar{\psi}) = V(\sigma; t) \), where \( \sigma = \bar{\psi} \psi \) represents the fermion bilinear.  
When solving the flow equation for the effective potential \( V_k(\sigma; t) \), the solution exhibits singularities (nonanalytic behavior), which signal the onset of spontaneous chiral symmetry breaking.  
To address this, we employed the weak formulation of the partial differential equation, which accommodates solutions with such singularities.

From the solution of the fermionic mass function $M(\sigma; t) = \partial_\sigma V(\sigma; t)$, we extracted the QCD phase diagram for two quark flavors.  
By comparing our results with those obtained from other continuum functional methods that include mesonic fluctuations, we found that both the critical temperature and the curvature of the $T_c(\mu_B)$ line near vanishing chemical potential yield reasonable agreement.  
However, the prediction of the chiral criticality deviates significantly from those approaches, which is partly in line with our expectations.

%%%%%%%%%%%%%%%%%%%%%%%%%%%%%%%%%%%%%%%%%%%%%%%%%%%%%%%%%%%%%%%%%%%%%%
\begin{acknowledgments}
We thank Fei Gao, Shinya Matsuzaki, and Jan M. Pawlowski for their helpful discussions.
We also thank Fabian Rennecke and Shi Yin for providing the fRG data on the running gauge coupling.
M.\,Y. also thanks all the other members in fQCD collaboration~\cite{fQCD:2025}.
%%%%%%
Y.\,G. and M.\,Y. acknowledge the Institute for Theoretical Physics, Heidelberg University, and Tokyo Women's Christian University for the very kind hospitality during their stay.
Y.\,G. further acknowledges the School of Frontier Sciences, Nanjing University, for the warm hospitality during his visit.
%%%%%%
\end{acknowledgments}

\onecolumngrid
%%%%%%%%%%%%%%%%%%%%%%%%%%%%%%%%%%%%%%%%%%%%%%%%%%%%%%%%%%%%%%%%%%%%%%
\appendix
\section{Wetterich equation with sharp regulator}
\label[appendix]{app:WetterichEquationWithSharpRegulator}

Starting with the IR-regulated Schwinger's functional
\begin{align}
    e^{W_k[J]} = \int \mathcal{D}\Phi \, e^{-S_{\rm bare} - \Delta S_k + J \cdot \Phi},
    \label{eq:IRSchwingerFunctional}
\end{align}
and the quadratic regulator term is
\begin{align}
    \Delta S_k = \int_x \Phi^T \cdot \mathcal{R}_k(\partial) \cdot \Phi.
    \label{eq:defineDeltaS}
\end{align}
$k \in [0,\Lambda]$ denotes the RG scale.
The general structure of the regulator $\mathcal{R}_k(p)$ in the momentum space is given by the following supermatrix form
\begin{align}
    \mathcal{R}_k(p) = \pmat{
    R_k^A & 0 & 0 \\
    0 & 0 & R_k^{\psi, T} \\
    0 & R_k^{\psi} & 0
    }.
\end{align}
Given the definition of the 1PI effective averaged action $\Gamma_k$ by 
\begin{align}
    \Gamma_k[\Phi] = \sup_J \left( J \cdot \Phi - W_k[J] \right) - \Delta S_k[\Phi],
    \label{eq:defineGammak}
\end{align}
where $W_k[J]$ denotes the Schwinger's functional and $\Delta S_k[\Phi]$ is the regulator term quadratic in fields,
%These two terms are defined in Eq.~\eqref{eq:IRSchwingerFunctional} and Eq.~\eqref{eq:defineDeltaS}.
we adopt the following conditions
\begin{align}
    \Gamma_k \rightarrow 
    \begin{cases}
        S_{\rm bare} & \quad k \rightarrow \infty, \\[1ex]
        \Gamma & \quad k \rightarrow 0.
    \end{cases}
\end{align}
This results in
\begin{align}
    \mathcal{R}_k \rightarrow 
    \begin{cases}
        \infty & \quad k \rightarrow \infty, \\[1ex]
        0 & \quad k \rightarrow 0.
    \end{cases}
\end{align}
After taking the derivative with respect to the RG scale $t$ on both sides of \cref{eq:IRSchwingerFunctional}, and performing the Legendre transformation, the flow equation for the 1PI effective action, i.e., the Wetterich equation \labelcref{eq:WetterichEq},
\begin{align}
    \partial_t \Gamma_k[\Phi] &= \frac{1}{2} \operatorname{STr} \left[ \left( \Gamma_k^{(2)} + \mathcal{R}_k \right)^{-1} \partial_t \mathcal{R}_k \right],
    \label{eq:appWetterichEq}
\end{align}
is found. See also \cite{Ellwanger:1993mw,Morris:1993qb}.
The UV boundary condition of \cref{eq:appWetterichEq} is the bare action $S_{\rm bare}$.

In this work, we apply the following spatial sharp regulator to regulate the action at finite temperature
\begin{align}
    \mathcal{R}_k({p}) = \mathcal{K}_k \, r_k(\bvec{p}),
\end{align}
where $r_k(\bvec p) = \frac{1}{\theta\left( |\bvec{p}| - k \right)} - 1$.
The tree-level kinetic matrix is given by
\begin{align}
    \mathcal{K}_k = \pmat{
    Z^A_k {p^2} \Pi_{\mu\nu}^{1/\xi} & 0 & 0 \\
    0 & 0 & Z^\psi_k i \slashed{{p}}^T \\
    0 & Z^\psi_k i \slashed{{p}} & 0
    }.
\end{align}
where $\Pi_{\mu\nu}^{1/\xi} = \Pi_{\mu\nu}^\perp + \frac{1}{\xi}\Pi_{\mu\nu}^\parallel$, and
\begin{align}
    \Pi_{\mu\nu}^\perp(p) = \delta_{\mu\nu} - \frac{p_\mu p_\nu}{p^2} , \qquad \Pi_{\mu\nu}^\parallel = \frac{p_\mu p_\nu}{p^2}.
    \label{eq:propAndParaProjection}
\end{align}
By differentiating the regulator with respect to the RG time $t$, we get
\begin{align}
    \partial_t \mathcal{R}_k = \mathcal{K}_k \frac{k\delta\left( |\bvec{p}| - k \right)}{\theta^2\left( |\bvec{p}| - k \right)},
\end{align}
where we have ignored the explicit anomalous dimension contribution from $\eta \sim - \partial_t \log Z$ in the loops.
Also, the Hessian matrix can be decomposed into
\begin{align}
    \Gamma_k^{(2)} = \mathcal{K}_k + \mathcal{V}_k.
\end{align}
By extracting the momentum integration of the supertrace in the flow equation, the flow equation is then deformed as
\begin{align}
    \frac{1}{2}\operatorname{STr} \left[ \left( \Gamma_k^{(2)} + \mathcal{R}_k \right)^{-1} \partial_t \mathcal{R}_k  \right] &= \frac{1}{2}T\sum_{n = -\infty}^{\infty} \int \frac{\df^3 \bvec{p}}{(2\pi)^3} \operatorname{str} \left[ \left( \mathcal{K}_k/\theta\left( |\bvec{p}| - k \right) + \mathcal{V}_k \right)^{-1} \mathcal{K}_k \frac{k\delta\left( |\bvec{p}| - k \right)}{\theta^2\left( |\bvec{p}| - k \right) } \right] \nonumber\\
    &= \frac{1}{2}T\sum_{n = -\infty}^{\infty} \int \frac{\df^3 \bvec{p}}{(2\pi)^3} {k\delta\left( |\bvec{p}| - k \right)}\operatorname{str} \left[ \theta^{-1}\left( |\bvec{p}| - k \right)\left( \mathcal{K}_k  + \mathcal{V}_k \cdot \theta\left( |\bvec{p}| - k \right) \right)^{-1} \mathcal{K}_k  \right]\nonumber\\
    &= \frac{1}{2}T\sum_{n = -\infty}^{\infty} \int \frac{\df^3 \bvec{p}}{(2\pi)^3} {k\delta\left( |\bvec{p}| - k \right)}\operatorname{str} \left[ \theta^{-1}\left( |\bvec{p}| - k \right) - \left( \mathcal{K}_k  + \mathcal{V}_k \cdot \theta\left( |\bvec{p}| - k \right) \right)^{-1} \mathcal{V}_k  \right].
    \label{eq:intermidiateToSharp}
\end{align}
The first term of the last line in \cref{eq:intermidiateToSharp} is a field-independent and divergent term, thus dropped there.
Following the \textit{Morris-Lemma}~\cite{Morris:1993qb}, we deform the second term as
\begin{align}
    &-\delta\left( |\bvec{p}| - k \right) \left( \mathcal{K}_k  + \mathcal{V}_k \cdot \theta\left( |\bvec{p}| - k \right) \right)^{-1} \mathcal{V}_k \nonumber\\
    &\quad= -\delta\left( |\bvec{p}| - k \right) \int_0^1 \df \theta \, \left( \mathcal{K}_k  + \mathcal{V}_k \cdot \theta \right)^{-1} \mathcal{V}_k \nonumber\\
    &\quad= -\delta\left( |\bvec{p}| - k \right) \log \left[ \frac{\mathcal{K}_k  + \mathcal{V}_k}{\mathcal{K}_k} \right].
    \label{eq:MorrisLemma}
\end{align}
Thus, the Wetterich equation is reduced to
\begin{align}
    \partial_t \Gamma_k[\Phi] = -\frac{1}{2}T\sum_{n = -\infty}^{\infty} \int \frac{\df^3 \bvec{p}}{(2\pi)^3} {k\delta\left( |\bvec{p}| - k \right)}\operatorname{str}\log \Gamma_k^{(2)}.
\end{align}
Hereafter, for simplicity, we introduce the following short-hand notation, replacing the integral/summation with
\begin{align}
    T\sum_{n = -\infty}^{\infty} \int \frac{\df^3 \bvec{p}}{(2\pi)^3}{k\delta\left( |\bvec{p}| - k \right)} \rightarrow \int_{p,\rm shell}.
    \label{eq:integral/Summation}
\end{align}

As the kernel of the flow equation, the Hessian has the following supermatrix form in QCD
\begin{align}
    \Gamma_{k,a,b}^{(2)} = \pmat{
    H_{BB} & H_{BF} \\
    H_{FB} & H_{FF}
    },
    \label{eq:generalHessian}
\end{align}
where $B$ and $F$ denote the bosonic and fermionic superfield components, respectively.

\begin{figure}
    \centering
    \includegraphics[width=0.5\linewidth]{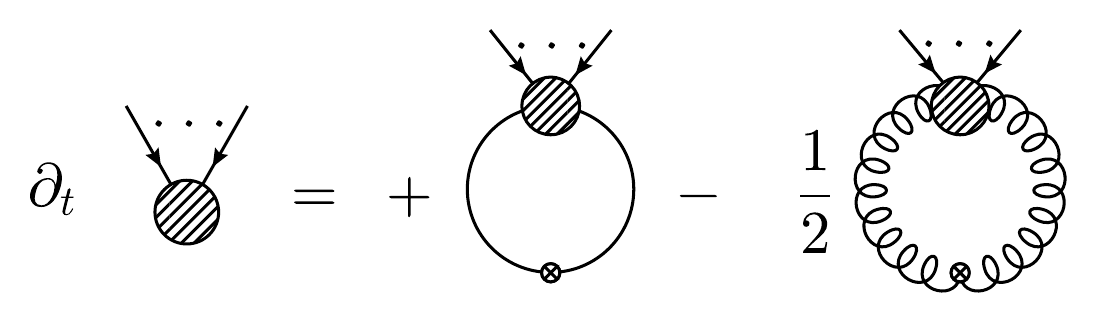}
    \caption{
    Diagrammatic expression of the flow equation \labelcref{eq:generalflowform}.
    The solid and wiggled lines denote the fermion and gluon propagators, respectively.
    The blobs on the propagator show the all-order external field dependence, denoted by the solid fermionic lines.
    The crossed circle denotes the regulator insertion, which implies the momentum shell integration.
    }
    \label{fig:FlowOfAction}
\end{figure}

In this work, we adopt the following form of the flow equation
\begin{align}
    \partial_t \Gamma_k
    &= -\frac{1}{2}\int_{p,\rm shell} \operatorname{str} \log \Gamma_{k,a,b}^{(2)} \nonumber\\
    &= \frac{1}{2} \int_{p,\rm shell} \biggl[\operatorname{tr}\log{H_{FF}} - \operatorname{tr}^\prime\log{(H_{BB} - H_{BF}H_{FF}^{-1}H_{FB})} \biggl],
    \label{eq:generalflowform}
\end{align}
where the prime on the trace indicates that the trace acts only on the Lorentz and adjoint color spaces.
The derivation of \cref{eq:generalflowform} can be found, e.g., in Appendix~B of the Ref.~\cite{Huang:2024ypj}.
This expression shows a intuitive diagrammatical interpretation, i.e., the flow equation with external fermion field is given by the one-loop of the fermion propagator $H_{FF}^{-1}$ and the one-loop of the gluonic propagator $H_{BB}^{-1}$ with the dressing of the effective self-energy $H_{BF}H_{FF}^{-1}H_{FB}$.
The corresponding Feynman diagram representation is shown in \Cref{fig:FlowOfAction}.

%

%%%%%%%%%%%%%%%%%%%%%%%%%%%%%%%%%%%%%%%%%%%%%%%%%%%%%%%%%%%%%%%%%%%%%%
\section{Flow equation of the fermionic potential \texorpdfstring{$V_k(\psi,\bar{\psi})$}{}}
\label[appendix]{app:DeriveFlowEq}

\subsection{Hessian}
\label[appendix]{app:DeriveHessian}

Given the truncated effective action \labelcref{eq:IREA}, the Hessian matrix is obtained by performing the second functional derivative with respect to the superfield component, like in \cref{eq:defineHessian}.
By using the notation defined in \cref{eq:generalHessian}, we write the bosonic and the fermionic components of the Hessian individually as
\begin{align}
    &H_{BB} = \delta_{p, q} Z^A_k (D^{-1})^{ab}_{\mu\nu}(p_A), \nonumber\\
    &H_{FF} = \pmat{
    \frac{\overset{\rightarrow}{\delta}}{\delta \psi^T(-p_\psi)} \mathcal{V}_k^\psi \frac{\overset{\leftarrow}{\delta}}{\delta \psi(q_\psi)} & \delta_{p, q} Z^\psi_k i \slashed{p}^{+,T}_\psi + \frac{\overset{\rightarrow}{\delta}}{\delta \psi^T(-p_\psi)} \mathcal{V}_k^\psi \frac{\overset{\leftarrow}{\delta}}{\delta \bar{\psi}^T(-q_\psi)} \\
    \delta_{p, q} Z^\psi_k i \slashed{p}^{-}_\psi + \frac{\overset{\rightarrow}{\delta}}{\delta \bar{\psi}(p_\psi)} \mathcal{V}_k^\psi \frac{\overset{\leftarrow}{\delta}}{\delta \psi(q_\psi)} & \frac{\overset{\rightarrow}{\delta}}{\delta \bar{\psi}(p_\psi)} \mathcal{V}_k^\psi \frac{\overset{\leftarrow}{\delta}}{\delta \bar{\psi}^T(-q_\psi)}
    },\nonumber\\
    &H_{BF} = \int_x \, e^{-i (p_A - q_\psi)\cdot x} \pmat{
        i g_s \bar{\psi}(x) \gamma_\mu T^a , & -i g_s  ( T^a \gamma_\mu \psi(x))^T
    },\nonumber\\
    &H_{FB} = \int_x \, e^{-i(p_\psi-q_A)\cdot x} \pmat{
        -i g_s (\bar{\psi}(x)\gamma_\nu T^b)^T \\
        i g_s \gamma_\nu T^b \psi(x)
    },
    \label{eq:generalHessianExplicit}
\end{align}
with $p_{A,\mu} = (2n\pi T, \bvec{p})$, $p_{\psi,\mu} = ((2n+1)\pi T, \bvec{p})$, $p_{\psi,\mu}^{\pm} = p_{\psi,\mu} \pm i \mu_q$, and
\begin{gather}
    (D^{-1})^{ab}_{\mu\nu}(p) = \frac{\delta^{ab}}{p^2} \Pi_{\mu\nu}^{1/\xi}(p) = \frac{\delta^{ab}}{p^2} \left[ \delta_{\mu\nu} - (1 - \frac{1}{\xi}) \frac{p_\mu p_\nu}{p^2} \right] \equiv \delta^{ab} D^{-1}_{\mu\nu}(p), \nonumber\\
    \mathcal{V}_k^\psi = -\int_x \, V_k(\psi,\bar{\psi}).
\end{gather}
According to the discussion around \cref{eq:scalarBilinearChannelOfVkExpansion}, only the scalar bilinear form within the LPA is taken into account in the fermionic potential, and we have the following properties
\begin{align}
    &\frac{\overset{\rightarrow}{\delta}}{\delta \psi^T(-p_\psi)} \sigma^n(x) \frac{\overset{\leftarrow}{\delta}}{\delta \psi(q_\psi)} = - n (n-1) \sigma^{n-2}(x) \bar{\psi}^T(x) \bar{\psi}(x) e^{-i(p_\psi - q_\psi)\cdot x}, \nonumber\\
    &\frac{\overset{\rightarrow}{\delta}}{\delta \bar{\psi}(p_\psi)} \sigma^n(x) \frac{\overset{\leftarrow}{\delta}}{\delta \bar{\psi}^T(-q_\psi)} = - n (n-1) \sigma^{n-2}(x) \psi(x) \psi^T(x) e^{-i(p_\psi - q_\psi)\cdot x}, \nonumber\\
    &\frac{\overset{\rightarrow}{\delta}}{\delta \psi^T(-p_\psi)} \sigma^n(x) \frac{\overset{\leftarrow}{\delta}}{\delta \bar{\psi}^T(-q_\psi)} = + n (n-1) \sigma^{n-2}(x) \bar{\psi}^T(x) \psi^T(x) e^{-i(p_\psi - q_\psi)\cdot x} - n \sigma^{n-1}(x) \textbf{1}_{\rm total} e^{-i(p_\psi - q_\psi)}, \nonumber\\
    &\frac{\overset{\rightarrow}{\delta}}{\delta \bar{\psi}(p_\psi)} \sigma^n(x) \frac{\overset{\leftarrow}{\delta}}{\delta \psi(q_\psi)} = + n (n-1) \sigma^{n-2}(x) \psi(x) \bar{\psi}(x) e^{-i(p_\psi - q_\psi)\cdot x} + n \sigma^{n-1}(x) \textbf{1}_{\rm total} e^{-i(p_\psi - q_\psi)}.
\end{align}
Following the standard large-$N$ expansion, we shall only keep the leading orders that are proportional to the singlet tensorial structure $\textbf{1}_{\rm total} = \textbf{1}_{\rm Dirac} \otimes \textbf{1}_{\rm flavor} \otimes \textbf{1}_{\rm color}$.
Thus the matrix elements of $H_{FF}$ reads
\begin{gather}
    \frac{\overset{\rightarrow}{\delta}}{\delta \psi^T(-p_\psi)} \mathcal{V}_k^\psi \frac{\overset{\leftarrow}{\delta}}{\delta \psi(q_\psi)} \rightarrow 0, \quad \frac{\overset{\rightarrow}{\delta}}{\delta \bar{\psi}(p_\psi)} \mathcal{V}_k^\psi \frac{\overset{\leftarrow}{\delta}}{\delta \bar{\psi}^T(-q_\psi)} \rightarrow 0, \nonumber\\
    \frac{\overset{\rightarrow}{\delta}}{\delta \psi^T(-p_\psi)} \mathcal{V}_k^\psi \frac{\overset{\leftarrow}{\delta}}{\delta \bar{\psi}^T(-q_\psi)} \rightarrow + \textbf{1}_{\rm total} \int_x \, e^{-i(p_\psi - q_\psi)} \partial_\sigma V(\sigma),\quad
    \frac{\overset{\rightarrow}{\delta}}{\delta \bar{\psi}(p_\psi)} \mathcal{V}_k^\psi \frac{\overset{\leftarrow}{\delta}}{\delta \psi(q_\psi)} \rightarrow - \textbf{1}_{\rm total} \int_x \, e^{-i(p_\psi - q_\psi)} \partial_\sigma V(\sigma).
    \label{eq:secondFermionDerivativeLargeN}
\end{gather}

Within the LPA' framework, the external background fields are set to be homogeneous to project the flow onto the vanishing external momentum. 
However, fermions exceed a minimum thermal excitation in the temporal direction due to their anti-symmetric nature during the period $x_0 \in [0,1/T)$.
Thus, we take the following external field backgrounds, which keep this feature
\begin{align}
    A_\mu^a = 0, \qquad \psi(x) = e^{i \pi T x_0} \psi, \qquad \bar{\psi}(x) = e^{-i \pi T x_0} \bar{\psi}.
    \label{eq:externalFieldBackgroundChoice}
\end{align}

Inserting \cref{eq:externalFieldBackgroundChoice} into \cref{eq:secondFermionDerivativeLargeN,eq:generalHessian}, and performing the spacetime integral/summation, we obtain the Hessian matrix under the current truncation scheme as
\begin{align}
    &H_{BB} = \delta_{p_A, q_A} Z^A_k (D^{-1})^{ab}_{\mu\nu}(p_A), \nonumber\\
    &H_{FF} = \delta_{p_\psi, q_\psi} \pmat{
   0 & Z^\psi_k i \slashed{p}^{+,T}_\psi + \partial_\sigma V \\
    Z^\psi_k i \slashed{p}^{-}_\psi - \partial_\sigma V & 0
    },\nonumber\\
    &H_{BF} = \pmat{\delta_{p_A,q_\psi}^{-}
        i g_s \bar{\psi} \gamma_\mu T^a , & \delta_{p_A,q_\psi}^{+}(-i g_s)  ( T^a \gamma_\mu \psi(x))^T
    },\nonumber\\
    &H_{FB} = \pmat{
        \delta_{q_A,p_\psi}^{+}(-i g_s) (\bar{\psi}(x)\gamma_\nu T^b)^T \\
        \delta_{q_A,p_\psi}^{-} i g_s \gamma_\nu(x) T^b \psi
    },
    \label{eq:LPAPrimeHessian}
\end{align}
where
\begin{align}
    \delta_{p_{A/\psi}, q_{A/\psi}} = (2\pi)^3 \delta^{(3)}(\bvec{p} - \bvec{q}) \delta_{n_p,n_q}/T, \qquad \delta_{p_A,q_\psi}^{\pm} = \begin{cases}
        (2\pi)^3 \delta^{(3)}(\bvec{p} - \bvec{q}) \delta_{n_p,n_q}/T, & \quad \text{for $+$,} \\[1ex]
        (2\pi)^3 \delta^{(3)}(\bvec{p} - \bvec{q}) \delta_{n_p,n_q+1}/T, & \quad \text{for $-$}.
    \end{cases}
\end{align}
Here, we introduce an alternative notation for $H_{FF}$ such that
\begin{align}
    H_{FF} = \delta_{p_\psi, q_\psi} \pmat{
    0 & S^{(T),-1}(p_\psi^+) \\
    S^{-1}(p_\psi^-) & 0
    },
    \label{eq:inverseQuarkPropagators}
\end{align}
and its inverse is obtained as
\begin{align}
    H_{FF}^{-1} = \delta_{p_\psi, q_\psi} \pmat{
    0 & S(p_\psi^-) \\
    S^{(T)}(p_\psi^+) & 0
    } = \delta_{p_\psi, q_\psi} \pmat{
    0  & \frac{-i \slashed{p}^-_\psi - \partial_\sigma V}{(p_{\psi,0} - i\mu_q)^2 + \bvec{p}^2 + (\partial_\sigma V)^2} \\
       \frac{-i \slashed{p}^{+,T}_\psi + \partial_\sigma V}{(p_{\psi,0} + i\mu_q)^2 + \bvec{p}^2 + (\partial_\sigma V)^2}  & 0
    }.
\end{align}

\subsection{Derivation of the flow equation}

With the truncated Hessian \labelcref{eq:LPAPrimeHessian} at hand, we now start to derive the closed form of the flow equation for the fermionic potential.
First, we focus on the right-hand side of the flow equation \labelcref{eq:generalflowform}.
The fermionic loop reads
\begin{align}
    \frac{1}{2} \int_{p,\rm shell} \operatorname{tr}\log{H_{FF}} = \frac{\Omega_4}{2} \int_{p, \rm shell} \biggl[ \operatorname{tr}\log{S^{-1}(p^-_\psi)} + \operatorname{tr}\log({S^{(T)}(p^+_\psi))^{-1}} \biggl],
    \label{eq:fermionicLoop}
\end{align}
where $\Omega_4 = \beta\mathcal {V}_3$ denotes the spacetime volume, and the trace is taken in the spinor, flavor, and color spaces.
The bosonic loop is obtained as
\begin{align}
    & \frac{1}{2} \int_{p,\rm shell} \operatorname{tr}^\prime\log{(H_{BB} - H_{BF}H_{FF}^{-1}H_{FB})} \nn
    &\quad = \frac{\Omega_4}{2} \int_{p, \rm shell} \operatorname{tr}^\prime\log \biggl\{ (D^{-1})^{ab}_{\mu\nu}(p_A) + g_s^2 \bar{\psi} \gamma_\mu T^a S(p^-_\psi) \gamma_\nu T^b \psi + g_s^2 ( T^a \gamma_\mu \psi)^T S^{(T)}(p^+_\psi) (\bar{\psi}\gamma_\nu T^b)^T  \biggl\} \nn
    &\quad = \frac{\Omega_4}{2} \int_{p,\rm shell} \operatorname{tr}^\prime\log \biggl\{ (D^{-1})^{ab}_{\mu\nu}(p_A) + g_s^2 \bar{\psi} \gamma_\mu T^a S(p^-_\psi) \gamma_\nu T^b \psi + g_s^2 \bar{\psi} \gamma_\nu T^b S(-p^+_\psi) \gamma_\mu T^a \psi \biggl\}
    \label{eq:bosonicLoop}
\end{align}

In the flow equation~\labelcref{eq:flowinsharpcutoff}, we have omitted the bare kinetic term contributions, i.e., the denominator in the logarithm of \cref{eq:MorrisLemma}.
These terms read
\begin{align}
    \frac{1}{2} \int_{p,\rm shell} \operatorname{str}\log{\mathcal{K}_k^{-1}} = +\frac{\Omega_4}{2} \int_{p, \rm shell} \biggl[ \operatorname{tr}\log{S_0^{-1}(p^-_\psi)} + \operatorname{tr}\log{(S^{(T)}(p^+_\psi))^{-1}} \biggl] + \frac{1}{2} \int_{p,\rm shell}  \operatorname{tr}^\prime\log (D)^{ab}_{\mu\nu}(p_A).
    \label{eq:subtructionTerms}
\end{align}

Performing the field strength renormalization $A_\mu^a \rightarrow  A_\mu^a/\sqrt{Z^{A}_k}$, $\psi \rightarrow  \psi/\sqrt{Z^{\psi}_k}$ and $\bar\psi \rightarrow  \bar\psi/\sqrt{Z^{\psi}_k}$, the gauge coupling is replaced by the renormalized one, and the left-hand side of the flow equation \labelcref{eq:flowinsharpcutoff} becomes
\begin{align}
    \partial_t \Gamma_k = - \Omega_4 \left( \partial_t V_k - \eta_\psi \sigma \partial_\sigma V_k \right) + \cdots,
    \label{eq:LHSoftheflow}
\end{align}
where $\eta_\psi =- \partial_t \log Z^\psi_k$ is the fermion anomalous dimension, and will be derived in \Cref{App:QuarkAnomalousDimension}.

Combining \cref{eq:fermionicLoop} to \cref{eq:LHSoftheflow}, the flow equation of the fermionic potential is then obtained as
\begin{align}
    \partial_t V_k = &\eta_\psi \sigma \partial_\sigma V_k - \frac{1}{2} \int_{p, \rm shell} \operatorname{tr}\biggl[ \log{S^{-1}(p^-_\psi)} + \log{(S^{(T)}(p^+_\psi))^{-1}} \biggl] + \frac{1}{2} \int_{p,\rm shell} \operatorname{tr}^\prime \log \biggl\{ \delta^{ab} \delta_{\mu\nu} + \mathcal{A}^{ab}_{\mu\nu} + \mathcal{B}^{ab}_{\mu\nu} \biggl\} \nonumber\\
    & + \frac{1}{2} \int_{p, \rm shell} \operatorname{tr} \biggl[ \log{S_0^{-1}(p^-_\psi)} + \log{(S^{(T)}_0(p^+_\psi))^{-1}} \biggl],
    \label{eq:flowOfVkInGeneral}
\end{align}
in which we have defined
\begin{align}
    &\mathcal{A}^{ab}_{\mu\nu} \equiv g_s^2 \bar{\psi} \gamma_\mu T^a S(p^-_\psi) \gamma_\rho D_{\rho\nu}(p_A) T^b \psi,
    \label{eq:defineAandBtensors1}
    \\
    &\mathcal{B}^{ab}_{\mu\nu} \equiv g_s^2 \bar{\psi} T^b D_{\rho\nu}(p_A) \gamma_\rho S(-p^+_\psi) \gamma_\mu  T^a \psi.
    \label{eq:defineAandBtensors}
\end{align}
This is the explicit form of \cref{eq:flowOfVkInGeneralMaintext}.

\subsection{Ladder and non-ladder Approximations}
\label[appendix]{sec:LadderNonladder}

Although the flow equations with the same truncation and approximation schemes are derived in the vacuum case in Ref.~\cite{Aoki:2012mj}, there are still subtleties at finite temperature and quark chemical potential.
Thus, a detailed derivation from the beginning is necessary.

\subsubsection{``Ladder'' case}

To close the flow equation, we compute the gluonic trace in \cref{eq:flowOfVkInGeneral}.
We expand the logarithm in terms of the polynomial of $\mathcal A$ and $\mathcal B$
\begin{align}
    \operatorname{tr}^\prime \log( 1 + \mathcal{A} + \mathcal{B}) = - \sum_{n=1}^\infty \frac{(-1)^n}{n} \operatorname{tr}^\prime \left( \mathcal{A} + \mathcal{B} \right)^n.
    \label{eq: vertex expansion}
\end{align}
This is essentially the vertex expansion of the bilinear quantities $\bar{\psi} \mathcal{O} \psi$.
The $n$-th order of the polynomial corresponds to the one-loop diagram with $n$-external background fields, and the order of contraction of $\mathcal A$ and $\mathcal B$ represents how this loop is constructed.
We then take the following ladder approximation that the vertices are contracted in a certain order, i.e., we only consider the $\mathcal A^n$ and the $\mathcal B^n$ terms and ignore other terms such as $\mathcal A\mathcal B\mathcal A^{n-2}$ in the current approximation scheme.
See also the discussion around \cref{eq:ladderAproxMainText}.
After that, the trace becomes
\begin{align}
    \operatorname{tr}^\prime \log( 1 + \mathcal A + \mathcal B) \quad \overset{\rm Ladder}{\longrightarrow} \quad - \sum_{n=1}^\infty \frac{(-1)^n}{n} \operatorname{tr}^\prime \left( \mathcal A^n +\mathcal B^n \right).
\end{align}

Next, we project the operator into the subspace of the scalar coupling channel using the generalized Fierz transformation \cite{Aoki:2012mj}
\begin{align}
    \bar{\psi}_1 \mathcal{O}_1 \psi_2 \bar{\psi}_2 \mathcal{O}_2 \psi_3 \cdots \bar{\psi}_n \mathcal{O}_n \psi_1 \quad \rightarrow \quad (-1)^{n+1} F_n \prod_{i=1}^n \bar{\psi}_i \textbf{1}_{\rm spinor} \lambda^0 T^0 \psi_i,
\end{align}
where $T^0 \equiv \textbf{1}_{\rm color}/\sqrt{2 N_c}$ is the normalized identity in the color space and $\lambda^0 \equiv \textbf{1}_{\rm flavor}/\sqrt{2 N_f}$ is the normalized identity in the flavor space.
The Fierz coefficient reads
\begin{align}
    F_n = \operatorname{tr} \biggl[ \textbf{1}_{\rm spinor} \lambda^0 T^0 \mathcal{O}_1 \cdots \textbf{1}_{\rm spinor} \lambda^0 T^0 \mathcal{O}_n \biggl],
\end{align}
where the trace is taken in the spinor, the flavor space, and the color space.
The Fierz coefficients of $\mathcal A^n$ and $\mathcal B^n$ are obtained by
\begin{align}
    F^{\mathcal A}_n &= \left( \frac{g_s^{2}}{4 N_c N_f} \right)^n \operatorname{tr} \biggl[ \gamma_{\mu_n} T^{a_n} S(p^-_\psi) \gamma_{\mu_1} D_{\mu_1 \mu_2}(p_A) T^{a_1} \gamma_{\mu_1} T^{a_1} \cdots S(p^-_\psi) \gamma_{\mu_n} D_{\mu_n \mu_1}(p_A) T^{a_n} \biggl] \nn
    &=\left( \frac{g_s^{2} C_2}{4 N_c N_f} \right)^n \operatorname{tr} \biggl[ S(p^-_\psi) \gamma_\mu D_{\mu \nu}(p_A)\gamma_\nu \biggl]^n \nn
    &= \left( \frac{g_s^{2} C_2}{4 N_c N_f} \right)^n \operatorname{tr} \biggl[ S(p^-_\psi) \frac{(3+\xi)}{p_A^2} \biggl]^n, \\[2ex]
    F^{\mathcal B}_n &= \left( \frac{g_s^{2} C_2}{4 N_c N_f} \right)^n \operatorname{tr} \biggl[ S(-p^+_\psi) \frac{(3+\xi)}{p_A^2} \biggl]^n,
\end{align}
where $T^a T^a = (N_c^2 - 1)/(2 N_c) \mathbf{1}_{\rm color} \equiv C_2 \mathbf{1}_{\rm color}$ is the Casimir operator of the fundamental representation of $SU(N_c)$.
Then the original operators are replaced by the projected ones
\begin{align}
    &- \sum_{n=1}^\infty \frac{(-1)^n}{n} \operatorname{tr}^\prime \left( \mathcal A^n +\mathcal B^n \right) \nn
    & \quad \overset{\rm F.T.}{\rightarrow} - \sum_{n=1}^\infty \frac{(-1)^{n+1}}{n} \left( \frac{g_s^{2} C_2}{4 N_c N_f} \right)^n \Biggl( \operatorname{tr} \biggl[ - S(p^-_\psi) \frac{(3+\xi)}{p_A^2} \biggl]^n + \operatorname{tr} \biggl[ - S(-p^+_\psi) \frac{(3+\xi)}{p_A^2} \biggl]^n  \Biggl) \nn
    & \quad = - \operatorname{tr} \log \biggl( 1 - C_2 g_s^2 \frac{3 + \xi}{4 p_A^2} S(p^-_\psi) \frac{\sigma}{N_c N_f}\biggl) - \operatorname{tr} \log \biggl( 1 - C_2 g_s^2 \frac{3 + \xi}{4 p_A^2} S(-p^+_\psi) \frac{\sigma}{N_c N_f}\biggl),
\end{align}
where F.T. denotes the Fierz transformation.

Combining with the fermionic loop and the left-hand side of the flow equation we derived, it turns out that
\begin{align}
    \partial_t V(\sigma ; t) =  \eta_\psi \sigma \partial_\sigma V &- \frac{1}{2} \int_{p,\rm shell} \operatorname{tr} \log \biggl( S^{-1}(p_\psi^-) - C_2 g_s^2 \frac{3 + \xi}{4 p_A^2}\sigma\biggl) \nn
    &- \frac{1}{2} \int_{p, \rm shell} \operatorname{tr} \log \biggl( - S^{-1}(p_\psi^+) + C_2 g_s^2 \frac{3 + \xi}{4 p_A^2} \sigma\biggl).
\end{align}
where we rescaled the $V$ and $\sigma$ by the factor $N_c N_f$
\begin{align}
    \sigma \rightarrow N_c N_f \sigma , \qquad V \rightarrow N_c N_f V,
    \label{eq: rescaling}
\end{align}
with $\partial_\sigma V$ unchanged.
After taking the trace and integrating out the loop momentum, we obtain the flow equation of the fermionic potential within the ladder approximation \labelcref{eq:ladderFlowMainText}
\begin{align}
    \partial_t V(\sigma ; t) =  \eta_\psi \sigma \partial_\sigma V &- T\sum_n  \frac{k^3}{2\pi^2} \log\left( \omega_{\psi,n}^2 + ( \sqrt{k^2+\mathcal{M}^2} - \mu_q )^2 \right)
    - T\sum_n  \frac{k^3}{2\pi^2} \log\left( \omega_{\psi,n}^2 + ( \sqrt{k^2+\mathcal{M}^2} + \mu_q )^2 \right),
    \label{eq: ladder flow}
\end{align}
where 
\begin{align}
    \mathcal{M} \equiv \partial_\sigma V + C_2 g_s^2 \frac{3 + \xi}{4 (\omega_{A,n}^2 + k^2)} \sigma,
    \label{eq: B definition in app}
\end{align}
and
\begin{align}
    \omega_{\psi,n} = (2n+1)\pi T , \qquad \omega_{A,n} = 2n \pi T.
\end{align}
Notice that the separation of the $\pm\mu_q$ terms is physically clear, that each term individually corresponds to the fermion or the anti-fermion loop that gains the opposite contribution from the quark chemical potential.
This is the effect induced by the ladder approximation that we do not incorporate with the mixing between fermion and anti-fermion, i.e., the broken ladder.

\subsubsection{``Non-ladder'' case}

Next, we go beyond the ladder approximation scheme by encountering the broken-ladder effect.
First, we consider the mixed quark-gluon vertex at vanishing quark chemical potential
\begin{align}
    (\mathcal{A}+\mathcal{B})^{ab}_{\mu\nu} &= g_s^2 \bar{\psi} \biggl[ T^a \gamma_\mu S(p_\psi) \gamma_\rho D_{\rho\nu}(p_A) T^b + T^b D_{\rho\nu}(p_A) \gamma_\rho S(-p_\psi) \gamma_\mu  T^a \biggl]\psi \nn
    &= \frac{g_s^2 D_{\rho\nu}(p_A)}{p_\psi^2 + (\partial_\sigma V)^2} \bar{\psi} \biggl[ T^a T^b \gamma_\mu \left(-i \slashed{p}_\psi - \partial_\sigma V\right) \gamma_\rho + T^a T^b \gamma_\rho \left(+i \slashed{p}_\psi - \partial_\sigma V\right) \gamma_\mu  \biggl]\psi 
    + (\text{$[T^a,T^b]$ term}).
    \label{eq:nonLadderApproximation}
\end{align}
To balance the color factors between different orders of combinations expanded from $(A+B)^n$.
This point motivates us to ignore the $[T^a, T^b]$ term, which is also discussed around \cref{eq:nonLadderApproximationMainText}.
Then we are able to deform the mixed quark-gluon by collecting the following expressions, where we define $M_\psi \equiv \partial_\sigma V$
\begin{align}
    (\text{$M_\psi$ term}) = - \frac{g_s^2}{p_\psi^2 + M_\psi^2} \bar{\psi} T^a T^b \biggl[ M_\psi \{\gamma_\mu,\gamma_\rho\}  D_{\rho\nu}(p_A) \biggl]\psi
    = - \frac{2 g_s^2}{p_\psi^2 + M_\psi^2} \bar{\psi} T^a T^b \biggl( M_\psi D_{\mu\nu}(p_A) \biggl) \psi,
    \label{eq:MixedQarkGluonVertexBeginningForMpsi}
\end{align}
\begin{align}
    (\text{$\slashed{p}_\psi$ term}) &= - \frac{g_s^2}{p_\psi^2 + M_\psi^2} \bar{\psi} T^a T^b \biggl( i p_{\psi,\sigma} \gamma_\mu \gamma_\sigma \gamma_\rho - i p_{\psi,\sigma} \gamma_\rho \gamma_\sigma \gamma_\mu \biggl)\frac{g_{\rho\nu}-(1-\xi)\frac{p_{A,\rho}p_{A,\nu}}{p_A^2}}{p_A^2}\psi \nn
    &= - \frac{g_s^2}{(p_\psi^2 + M_\psi^2)p_A^2} \bar{\psi} T^a T^b \biggl( i p_{\psi,\sigma} \gamma_\mu \gamma_\sigma \gamma_\nu - i p_{\psi,\sigma} \gamma_\nu \gamma_\sigma \gamma_\mu \biggl) \psi \nn
    &\quad + \frac{g_s^2 (1-\xi)}{(p_\psi^2 + M_\psi^2)p_A^4} \bar{\psi} T^a T^b \biggl( i p_{\psi,\sigma} p_{A,\rho} p_{A,\nu} \gamma_\mu \gamma_\sigma \gamma_\rho - i p_{\psi,\sigma} p_{A,\rho} p_{A,\nu} \gamma_\rho \gamma_\sigma \gamma_\mu \biggl) \psi.
    \label{eq:MixedQarkGluonVertexBeginningForpshashed}
\end{align}
For the last line in the $\slashed{p}_\psi$ term, we impose a further $O(4)$-approximation $\omega_{\psi}\approx \omega_{A}$ such that
\begin{align}
    \slashed{p}_A \slashed{p}_\psi \approx p_A \cdot p_\psi,
    \label{eq: O(4) limit}
\end{align}
which implies that this line vanishes under this approximation.
The approximation stated in \cref{eq: O(4) limit} might be spoiled in a large quark chemical potential and high temperature regime.
Using the following property of the $\gamma_5$ in the Euclidean spacetime
\begin{align}
    \gamma_\mu \gamma_\sigma \gamma_\nu = \delta_{\mu\sigma} \gamma_\nu + \delta_{\sigma\nu} \gamma_\mu - \delta_{\mu\nu} \gamma_\sigma + \epsilon_{\rho \mu\sigma \nu}  \gamma_5 \gamma_\rho,
\end{align}
we obtain
\begin{align}
    (\text{$\slashed{p}_\psi$ term}) &= - \frac{i g_s^2p_{\psi,\sigma}}{(p_\psi^2 + M_\psi^2)p_A^2} \bar{\psi} T^a T^b  \biggl( \epsilon_{\rho \mu\sigma \nu} \gamma_5 \gamma_\rho - \epsilon_{\rho \nu\sigma \mu} \gamma_5 \gamma_\rho  \biggl) \psi \nn
    &= - \frac{2 g_s^2}{(p_\psi^2 + M_\psi^2)p_A^2} \bar{\psi} T^a T^b  \biggl( i p_{\psi,\alpha} \epsilon_{\mu\nu\alpha\beta} \gamma_5 \gamma_\beta \biggl) \psi,
\end{align}
where we have changed the dummy indices $\sigma \rightarrow \alpha$ and $\rho \rightarrow \beta$.
The mixed quark-gluon vertex then reads
\begin{align}
    (\mathcal{A}+\mathcal{B})^{ab}_{\mu\nu} = - \frac{2 g_s^2}{p_\psi^2 + M_\psi^2} \bar{\psi} T^a T^b \biggl( i \frac{p_{\psi,\alpha}}{p_A^2} \epsilon_{\mu\nu\alpha\beta} \gamma_5 \gamma_\beta + M_\psi D_{\mu\nu}(p_A) \biggl) \psi.
\end{align}

For the $n$-th order term in \cref{eq: vertex expansion}, we obtain the following expression under the approximation scheme beyond the ladder
\begin{align}
     &\frac{(-1)^n}{2 n} \operatorname{tr}^\prime \left( \mathcal A+\mathcal B \right)^n \nn
     &\quad \overset{\rm B.L.}{\rightarrow} \frac{(-1)^n}{2 n} \left( - \frac{2 g_s^2}{p_\psi^2 + M_\psi^2} \right)^n \operatorname{tr}^\prime \biggl[  \bar{\psi} T^a T^b \biggl( i \frac{p_{\psi,\alpha}}{p_A^2} \epsilon_{\mu\nu\alpha\beta} \gamma_5 \gamma_\beta + M_\psi D_{\rho\nu}(p_A) \biggl) \psi \biggl]^n \nn
     &\quad \overset{\rm F.T.}{\rightarrow} \frac{(-1)^{n+1}}{2 n} \left( \frac{g_s^2 C_2}{2(p_\psi^2 + M_\psi^2)} \right)^n \operatorname{tr} \biggl[ i \frac{p_{\psi,\alpha}}{p_A^2} \epsilon_{\mu\nu\alpha\beta} \gamma_5 \gamma_\beta + M_\psi D_{\rho\nu}(p_A) \biggl]^n \left( \frac{\sigma}{N_c N_f}\right)^n,
     \label{eq:resultOfBalancingColorFactor}
\end{align}
where ``B.L.'' in the second line denotes the ``Beyond Ladder'' approximation scheme.
To evaluate this fermionic trace, we expand the matrix production $\biggl[ i \frac{p_{\psi,\alpha}}{p_A^2} \epsilon_{\mu\nu\alpha\beta} \gamma_5 \gamma_\beta + M_\psi D_{\rho\nu}(p_A) \biggl]^n$ and trace them individually.
First, we consider the $(M_\psi D)^n$ term
\begin{align}
    \operatorname{tr} \biggl[ M_\psi D_{\rho\nu}(p_A) \biggl]^n = \operatorname{tr} \biggl[ \frac{M_\psi}{p^2} \biggl( \Pi_{\mu\nu}^\perp(p_A) + \xi\Pi_{\mu\nu}^\parallel(p_A)  \biggl) \biggl]^n,
    \label{eq: pure Mpsi term}
\end{align}
where the transverse and longitudinal projection tensors are defined in \cref{eq:propAndParaProjection}.
Note here that 
\begin{align}
    \Pi_{\mu\rho}^\perp(p) \Pi_{\rho\nu}^\perp(p) = \Pi_{\mu\nu}^\perp(p) , \qquad \Pi_{\mu\rho}^\parallel(p) \Pi_{\rho\nu}^\parallel(p) = \Pi_{\mu\nu}^\parallel(p) , \qquad  \Pi_{\mu\rho}^\perp(p) \Pi_{\rho\nu}^\parallel(p) = 0.
\end{align}
\Cref{eq: pure Mpsi term} are deformed as
\begin{align}
    \operatorname{tr} \biggl[ M_\psi D_{\rho\nu}(p_A) \biggl]^n &= \left( \frac{M_\psi}{p_A^2} \right)^n \operatorname{tr} \biggl[ \Pi_{\mu\nu}^\perp(p_A) + \xi^n\Pi_{\mu\nu}^\parallel(p_A) \biggl] \nn
    &= 4 N_c N_f \left( \frac{M_\psi}{p_A^2} \right)^n (3+\xi^n).
\end{align}
Next, we consider the remaining terms.
We notice that the terms include the following expression
\begin{align}
    \frac{p_{A,\mu} p_{A,\nu}}{p_A^2} p_{\psi, \alpha} \epsilon_{\nu\rho\alpha\beta},
\end{align}
vanished in the $O(4)$-limit we mentioned around \cref{eq: O(4) limit} due to the anti-symmetric nature of the Levi-Civita symbol.
Thus, the only contributions to the mixed terms in the expansion of the matrix production come from the $\delta_{\mu\nu}$-structure of the gluon. 
We also notice that the terms with odd numbers of the $\epsilon$-terms vanish since they would contain odd numbers of $\gamma$-matrices.
Then, the expansion of the matrix production only includes the terms with even numbers of the $p_\alpha$.
Collecting all the ingredients we mentioned, the rest of the terms in the expansion of the matrix production read
\begin{align}
    \sum_{k=1}^{[\frac{n}{2}]} C_n^{2k} (-1)^k M^{n-2k} \operatorname{tr}\left[ \mathcal{T}^k \right],
\end{align}
where $[\frac{n}{2}]$ denotes the floor function of $\frac{n}{2}$, $C_n^{2k}$ is the combination number, and we have defined
\begin{align}
    \mathcal{T}_{\mu\nu} &\equiv p_{\psi,\alpha} p_{\psi,\alpha^\prime} \epsilon_{\mu\rho\alpha\beta}\epsilon_{\rho\nu\alpha^\prime\beta^\prime} \gamma_5 \gamma_{\beta}\gamma_5 \gamma_{\beta^\prime} \nn
    &= p^2_\psi \gamma_\mu\gamma_\nu + p_{\psi,\mu} \gamma_\nu \slashed{p}_\psi - p_{\psi,\nu} \gamma_\mu \slashed{p}_\psi + p^2_\psi \delta_{\mu\nu} - 2 p_{\psi,\mu} p_{\psi,\nu}.
\end{align}
Through algebraic calculations, we notice that
\begin{align}
    &\mathcal{T}_{\mu\rho} \mathcal{T}_{\rho\nu} = 5 p_\psi^2 \mathcal{T}_{\mu\nu} - 4 p_\psi^4 \Pi^\perp_{\mu\nu}(p_\psi) ,&
    &\mathcal{T}_{\mu\rho} \Pi^\perp_{\rho\nu}(p_\psi) = \mathcal{T}_{\mu\nu},&
    &\Pi^\perp_{\mu\rho}(p_\psi) \Pi^\perp_{\rho\nu}(p_\psi) = \Pi^\perp_{\mu\nu}(p_\psi),
    \label{eq:extremeNotification}
\end{align}
where the transverse projection tensor $\Pi^\perp_{\mu\nu}(p_\psi)$ is the same as the gluonic one.
With the observation of \cref{eq:extremeNotification}, we parameterize the structure of the $n$-th order of the tensor as
\begin{align}
    \left[ \mathcal{T}^k \right]_{\mu\nu} = p_\psi^{2k} \left( a_k \, \tilde{\mathcal{T}}_{\mu\nu} + b_k \, \Pi^\perp_{\mu\nu} \right),
\end{align}
where $\tilde{\mathcal{T}} \equiv \mathcal{T} / p^2_\psi$.
We then obtain the following recursion relations
\begin{align}
    a_{k+1} = 5 a_k + b_k , \qquad b_{k+1} = -4 b_k,
\end{align}
with the boundary conditions $a_1 = 1$ and $b_1 = 0$.
By solving these recursion relations, we obtain
\begin{align}
    a_k = \frac{1}{3}\left( 4^k - 1 \right) , \qquad b_k = -\frac{1}{3} \left( 4^k - 4 \right).
\end{align}
Combining with the traces
\begin{align}
    \operatorname{tr}\left[ \tilde{\mathcal{T}} \right] = 4 N_c N_f \times 6 , \quad \operatorname{tr}\left[ \Pi^\perp \right] = 4 N_c N_f \times 3,
\end{align}
we obtain
\begin{align}
    \operatorname{tr}\left[ \mathcal{T}^k \right] = 4 N_c N_f ( 2 \cdot 4^k-2 - 4^k + 4 ) p_\psi^{2k} = 4 N_c N_f (2 + 4^k) p_\psi^{2k}.
\end{align}

Combine all, the $n$-th order term is deformed into
\begin{align}
    \frac{(-1)^n}{2 n} \operatorname{tr}^\prime \left( \mathcal A+\mathcal B \right)^n \rightarrow  & 2 N_c N_f \frac{(-1)^{n+1}}{n} \left( \frac{g_s^2 C_2}{2(p_\psi^2 + M_\psi^2)p_A^2} \frac{\sigma}{N_c N_f}\right)^n \nn
    &\quad \times \biggl[ (\xi M_\psi)^n + \sum_{k=0}^{[\frac{n}{2}]} C_n^{2k} (-1)^k (2 + 4^k) M^{n-2k} p_\psi^{2k} \biggl].
\end{align}
Note that
\begin{align}
    \sum_{k=0}^{[\frac{n}{2}]} C_n^{2k} (-1)^k 2 M^{n-2k} p_\psi^{2k} &= \sum_{k^\prime=0}^{n} C_n^{2k}  M^{n-k^\prime} (-1)^{k^\prime} \sqrt{-p_\psi^2}^{k^\prime} + \sum_{k^\prime=0}^{n} C_n^{2k} M^{n-k^\prime} \sqrt{-p_\psi^2}^{k^\prime} \nn
    &= \left( M_\psi - \sqrt{-p_\psi^2} \right)^n + \left( M_\psi + \sqrt{-p_\psi^2} \right)^n, \nn[2ex]
    \sum_{k=0}^{[\frac{n}{2}]} C_n^{2k} (-1)^k 4^k M^{n-2k} p_\psi^{2k} &= \frac{1}{2}\left( M_\psi - 2 \sqrt{-p_\psi^2} \right)^n + \frac{1}{2}\left( M_\psi + 2 \sqrt{-p_\psi^2} \right)^n,
\end{align}
we are able to re-sum the vertex expansion back into the logarithmic form with $\tilde{\sigma} = \sigma/(N_c N_f)$
\begin{align}
   & \frac{1}{2} \operatorname{tr}^\prime \log(1+\mathcal A+\mathcal B) \nn
   & \rightarrow -N_c N_f \left\{ 2 \log \left[ 1 + \frac{g_s^2 C_2 \tilde{\sigma}}{2 p_A^2 (p_\psi^2 + M_\psi^2)} \left( M_\psi - \sqrt{-p_\psi^2} \right) \right] + 2 \log \left[ 1 + \frac{g_s^2 C_2 \tilde{\sigma}}{2 p_A^2 (p_\psi^2 + M_\psi^2)} \left( M_\psi + \sqrt{-p_\psi^2} \right) \right] \right. \nn
   & \quad \quad \quad \quad+ \log \left[ 1 + \frac{g_s^2 C_2 \tilde{\sigma}}{2 p_A^2 (p_\psi^2 + M_\psi^2)} \left( M_\psi - 2\sqrt{-p_\psi^2} \right) \right] + \log \left[ 1 + \frac{g_s^2 C_2 \tilde{\sigma}}{2 p_A^2 (p_\psi^2 + M_\psi^2)} \left( M_\psi + 2\sqrt{-p_\psi^2} \right) \right] \nn
   & \quad \quad \quad \quad+ \left. 2 \log \left[ 1 + \frac{g_s^2 C_2 \tilde{\sigma}}{2 p_A^2 (p_\psi^2 + M_\psi^2)} \left( \xi M_\psi \right) \right] \right\} \nn
   &= -N_c N_f \left\{ 2 \log \left[ 1 + \frac{g_s^2 C_2 \tilde{\sigma}}{p_A^2 (p_\psi^2 + M_\psi^2)} M_\psi + \left(\frac{g_s^2 C_2 \tilde{\sigma}}{2 p_A^2 (p_\psi^2 + M_\psi^2)}\right)^2 \left( M_\psi^2 + p_\psi^2 \right) \right] \right. \nn
   & \quad \quad \quad \quad + \log \left[ 1 + \frac{g_s^2 C_2 \tilde{\sigma}}{p_A^2 (p_\psi^2 + M_\psi^2)} M_\psi + \left(\frac{g_s^2 C_2 \tilde{\sigma}}{2 p_A^2 (p_\psi^2 + M_\psi^2)}\right)^2 \left( M_\psi^2 + 4p_\psi^2 \right) \right] \nn
   & \quad \quad \quad \quad+ \left. 2 \log \left[ 1 + \frac{g_s^2 C_2 \tilde{\sigma}}{2 p_A^2 (p_\psi^2 + M_\psi^2)} \left( \xi M_\psi \right) \right] \right\}.
\end{align}

For the fermionic loop, we obtain
\begin{align}
    \frac{1}{2}\operatorname{tr}\log{S^{-1}(p_\psi)} &= \frac{1}{4} \operatorname{tr}\log{S^{-1}(p_\psi)} + \frac{1}{4}\operatorname{tr}\log{\left(\gamma_5 S^{-1}(p_\psi) \gamma_5\right)} \nn
    &= \frac{1}{4}\operatorname{tr}\log (p_\psi^2 + M_\psi^2) \nn
    &= N_c N_f \log (p_\psi^2 + M_\psi^2), \\[1ex]
    \frac{1}{2}\operatorname{tr}\log{(S^{(T)}(p_\psi))^{-1}} &= N_c N_f \log (p_\psi^2 + M_\psi^2).
\end{align}
Combining all, the summation of the fermionic and gluonic loops reads
\begin{align}
    &\frac{1}{2}\operatorname{tr}\log{S^{-1}(p_\psi)} + \frac{1}{2}\operatorname{tr}\log{(S^{T}(p_\psi))^{-1}} - \frac{1}{2} \operatorname{tr}^\prime \log(1+\mathcal A+\mathcal B) \nn
    &= N_c N_f \left\{ 2 \log \left[ (p_\psi^2 + M_\psi^2) + \frac{g_s^2 C_2 \tilde{\sigma}}{p_A^2} M_\psi + \left(\frac{g_s^2 C_2 \tilde{\sigma}}{2 p_A^2 }\right)^2 \right] \right. \nn
    & \quad \quad \quad \quad + \log \left[ \frac{p_\psi^2 + M_\psi^2 + 2\frac{g_s^2 C_2 \tilde{\sigma}}{2 p_A^2} M_\psi + \left( \frac{g_s^2 C_2 \tilde{\sigma}}{2 p_A^2} \right)^2}{(p_\psi^2 + M_\psi^2)}  + \left(\frac{g_s^2 C_2 \tilde{\sigma}}{2 p_A^2 (p_\psi^2 + M_\psi^2)}\right)^2 3p_\psi^2 \right] \nn
    & \quad \quad \quad \quad+ \left. 2 \log \left[ 1 + \frac{g_s^2 C_2 \tilde{\sigma}}{2 p_A^2 (p_\psi^2 + M_\psi^2)} \left( \xi M_\psi \right) \right] \right\} \nn
    &= N_c N_f \left\{ 2 \log \left[ p_\psi^2 + \tilde{\mathcal M}^{\prime 2} \right] + \log \left[ \frac{p_\psi^2 + \tilde{\mathcal M}^{\prime 2}}{p_\psi^2 + M_\psi^2} + \frac{3 p_\psi^2 \tilde{\mathcal G}^2}{(p_\psi^2 + M_\psi^2)^2}  \right] + 2 \log \left[ 1 + \xi \frac{M_\psi\tilde{\mathcal G}}{p_\psi^2 + M_\psi^2} \right] \right\},
\end{align}
where we have defined
\begin{align}
    \tilde{\mathcal M}^{\prime} \equiv M_\psi +  C_2 \frac{g_s^2\tilde{\sigma}}{2 p_A^2} , \qquad \tilde{\mathcal G} \equiv C_2 \frac{g_s^2\tilde{\sigma}}{2 p_A^2}.
\end{align}
Then the flow equation of the fermionic potential can be obtained straightforwardly by taking the integral and summation \labelcref{eq:integral/Summation}
\begin{align}\label{eq: non-ladder flow}
    \partial_t V (\sigma;t) &= \eta_\psi \sigma \partial_\sigma V - T\sum_n \frac{k^3}{\pi^2} \log\left( \omega_{\psi,n}^2 + k^2 + {\mathcal M}^{\prime 2} \right) \nn
    &\quad - T\sum_n \frac{k^3}{2\pi^2} \log\left( \frac{\omega_{\psi,n}^2 + k^2 + {\mathcal M}^{\prime 2}}{ \omega_{\psi,n}^2 + k^2 + M_\psi^2} + \frac{3(\omega_{\psi,n}^2 + k^2) {\mathcal G}^2}{ (\omega_{\psi,n}^2 + k^2 + M_\psi^2)^2} \right) \nn
    &\quad - T\sum_n \frac{k^3}{\pi^2} \log\left( 1 + \xi \frac{M_\psi {\mathcal G}}{\omega_{\psi,n}^2 + k^2 + M_\psi^2} \right),
\end{align}
after we rescale the potential and the composite operator by \cref{eq: rescaling}, where
\begin{align}
    {\mathcal M}^{\prime} \equiv M_\psi + C_2 \frac{g_s^2 \sigma}{2 (\omega_{A,n}^2 + k^2)}, \qquad {\mathcal G} \equiv C_2 \frac{g_s^2 \sigma}{2 (\omega_{A,n}^2 + k^2)}.
\end{align}

For the case with finite quark chemical potential, \cref{eq:nonLadderApproximation} reads
\begin{align}
    \biggl(\mathcal A(\mu_q) + \mathcal B(\mu_q)\biggl)^{ab}_{\mu\nu} &= g_s^2 \bar{\psi} \biggl[ T^a \gamma_\mu S(p_\psi^{-}) \gamma_\rho D_{\rho\nu}(p_A) T^b + T^b D_{\rho\nu}(p_A) \gamma_\rho S(-p_\psi^{+}) \gamma_\mu  T^a \biggl]\psi.
\end{align}
Introducing the quark chemical potential brings in the asymmetry of the mixed quark-gluon vertex with respect to $\mu_q \rightarrow - \mu_q$.
To deal with it, we decompose the full vertex into symmetric and anti-symmetric parts of $\mu_q$:
\begin{align}
    \biggl(\mathcal A(\mu_q)+\mathcal B(\mu_q)\biggl)^{ab}_{\mu\nu} =& \frac{1}{2}\biggl(\mathcal A(\mu_q)+\mathcal B(-\mu_q)\biggl)^{ab}_{\mu\nu} + \frac{1}{2}\biggl(\mathcal A(-\mu_q)+\mathcal B(\mu_q)\biggl)^{ab}_{\mu\nu} \nonumber\\
    &+ \frac{1}{2}\left( \mathcal A^{ab}_{\mu\nu}(\mu_q) - \mathcal A^{ab}_{\mu\nu}(-\mu_q)\right) + \frac{1}{2}\left( \mathcal B^{ab}_{\mu\nu}(\mu_q) - \mathcal B^{ab}_{\mu\nu}(-\mu_q)\right).
    \label{eq:decompositionOfTheQuarkGluonVertex}
\end{align}
The formulation of the first line of \cref{eq:decompositionOfTheQuarkGluonVertex} is almost the same as what we have done before.
The first term of the second line is evaluated as
\begin{align}
    \frac{1}{2}\left( \mathcal A^{ab}_{\mu\nu}(\mu_q) - \mathcal A^{ab}_{\mu\nu}(-\mu_q)\right) &= \frac{1}{2}g_s^2 \bar{\psi} \biggl\{ T^a \gamma_\mu \left[ S(p_\psi^{-}) - S(p_\psi^{+}) \right] \gamma_\rho D_{\rho\nu}(p_A) T^b \biggl\}\psi \nonumber\\
    &= \frac{1}{2}g_s^2 \bar{\psi} \biggl\{ T^a \gamma_\mu \left[ \frac{(-i \slashed{p}_\psi - M_\psi)(p_\psi^{+,2} - p_\psi^{-,2}) - \mu_q \gamma_0 (p_\psi^{+,2} + p_\psi^{-,2} + 2 M_\psi^2)}{(p_\psi^{-,2} + M_\psi^2)(p_\psi^{+,2} + M_\psi^2)} \right] \gamma_\rho D_{\rho\nu}(p_A) T^b \biggl\} \psi.
\end{align}
Since the loop integral/summation traces over the Matsubara frequency from $-\infty$ to $\infty$, the combination $((p_\psi^{+})^2 - (p_\psi^{-})^2)$ would vanish except for the term with $p_{\psi,0}$.
This results in
\begin{align}
    \frac{1}{2}\left( \mathcal A^{ab}_{\mu\nu}(\mu_q) - \mathcal A^{ab}_{\mu\nu}(-\mu_q)\right) &= \frac{1}{2}g_s^2 \bar{\psi} \biggl\{ T^a \gamma_\mu \left[ \frac{-i p_{\psi,0} \gamma_0 ( 4i \mu_q p_{\psi,0} )- \mu_q \gamma_0 (p_\psi^{+,2} + p_\psi^{-,2} + 2 M_\psi^2)}{(p_\psi^{-,2} + M_\psi^2)(p_\psi^{+,2} + M_\psi^2)} \right] \gamma_\rho D_{\rho\nu}(p_A) T^b \biggl\} \psi \nonumber\\
    &=\frac{1}{2}g_s^2 \bar{\psi} \biggl\{ T^a \gamma_\mu \gamma_0 \left[ \frac{ 2 \mu_q (p_{\psi,0}^2 - |\bvec{p}|^{2} - M_\psi^2 + \mu_q^2)}{(p_\psi^{-,2} + M_\psi^2)(p_\psi^{+,2} + M_\psi^2)} \right] \gamma_\rho D_{\rho\nu}(p_A) T^b \biggl\} \psi,
    \label{eq:antiSymmetricA}
\end{align}
and
\begin{align}
    \frac{1}{2}\left( \mathcal B^{ab}_{\mu\nu}(\mu_q) - \mathcal B^{ab}_{\mu\nu}(-\mu_q)\right) = \frac{1}{2}g_s^2 \bar{\psi} \biggl\{ T^b D_{\rho\nu}(p_A) \gamma_\rho \left[ \frac{ 2 \mu_q (p_{\psi,0}^2 - |\bvec{p}|^{2} - M_\psi^2 + \mu_q^2)}{(p_\psi^{-,2} + M_\psi^2)(p_\psi^{+,2} + M_\psi^2)} \right]\gamma_0 \gamma_\mu  T^a  \biggl\} \psi.
    \label{eq:antiSymmetricB}
\end{align}
We find that \cref{eq:antiSymmetricA,eq:antiSymmetricB} show a different tensorial structure compared with \cref{eq:MixedQarkGluonVertexBeginningForpshashed}, and that the second line of \cref{eq:MixedQarkGluonVertexBeginningForpshashed} survives.
In fact, the anti-symmetric part of the mixed quark-gluon vertex brings in the imaginary part of it.
In this work, we do not deal with the complex flow (see, e.g., Ref.~\cite{Ihssen:2022xjv}), which means that we will project the flow onto the real part.
In this sense, we drop the anti-symmetric part as an approximation, which serves as a milder systematic error with the factor
\begin{align}
    f = \frac{(p_{\psi,0}^2 - |\bvec{p}|^{2} - M_\psi^2 + \mu_q^2)}{(p_\psi^{\pm,2} + M_\psi^2)},
\end{align}
correcting for the quark chemical potential.
This term can also be dropped in the Feynman gauge $\xi = 1$, alternatively.

In conclusion, the flow equation in the non-ladder case is summarized as
\begin{align}
    \partial_t V (\sigma;t) &= \eta_\psi \sigma \partial_\sigma V - \frac{1}{2} \left[ {\rm Flow}^{{\rm N.L.},-}_k(\sigma, M_\psi) + {\rm Flow}^{{\rm N.L.},+}_k(\sigma, M_\psi) \right],
    \label{eq:nonLadderFlowWithmu}
\end{align}
where the flow kernels are given by
\begin{align}
    {\rm Flow}^{{\rm N.L.},\pm}_k(\sigma, M_\psi) &= T\sum_n \frac{k^3}{\pi^2} \log\left( \omega_{\psi,n}^{\pm,2} + k^2 + {\mathcal M}^{\prime 2} \right) \nn
    &\quad + T\sum_n \frac{k^3}{2\pi^2} \log\left( \frac{\omega_{\psi,n}^{\pm,2} + k^2 + {\mathcal M}^{\prime 2}}{ \omega_{\psi,n}^{\pm,2} + k^2 + M_\psi^2} + \frac{3(\omega_{\psi,n}^{\pm,2} + k^2) {\mathcal G}^2}{ (\omega_{\psi,n}^{\pm,2} + k^2 + M_\psi^2)^2} \right) \nn
    &\quad + T\sum_n \frac{k^3}{\pi^2} \log\left( 1 + \xi \frac{M_\psi {\mathcal G}}{\omega_{\psi,n}^{\pm,2} + k^2 + M_\psi^2} \right),
    \label{eq:defineFlowName}
\end{align}
with $\omega_{\psi,n}^{\pm} = \omega_{\psi,n} \pm i \mu_q = (2n+1)\pi T \pm i \mu_q$.

With those approximations, the flow equation for the fermionic potential closed as a non-linear first-order partial differential equation.

\subsection{Quark anomalous dimension}
\label[appendix]{App:QuarkAnomalousDimension}

\begin{figure}[h]
    \centering
    \includegraphics[width=0.5\linewidth]{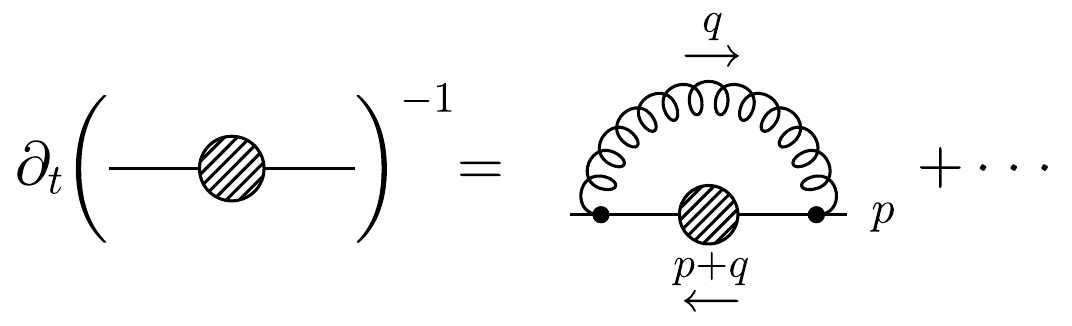}
    \caption{Diagrammatic expression of the flow equation of the quark two-point function, in which we only take into account the relevant part to evaluate the quark anomalous dimension in the current truncation scheme.
    The definitions of the lines are the same as in \Cref{fig:FlowOfAction}. }
    \label{fig:SunsetOfTwoPoint}
\end{figure}

To extract the flow of the two-point function with external fluctuations, we perform a formal vertex expansion of the gluonic loop with respect to the effective vertex $M_{BF} M_{FF}^{-1} M_{FB}$, carrying the external field dependence.
Since the fermionic one-loop, which results in the tadpole diagram, does not contribute to the anomalous dimension, we shall omit it in our formulation.
Also, we ignore the derivatives of the wave function with respect to the fields in the anomalous dimension, such as $Z^{(n)}(\sigma) = \partial_\sigma^n Z(\sigma)$, as an approximation, which stresses that the potential term would also be dominant in the flow of the wave function.
We then obtain the flow of the fermionic quadratic term at the vanished quark chemical potential as
\begin{align}
    &\int_{x,y,p} \bar{\psi}(x) \biggl(\partial_t \Gamma^{(2)}_{\bar{\psi}\psi}(\psi,\bar{\psi};p) \biggl)\psi(y) \, e^{-ip\cdot(x-y)} \nonumber\\
    &\quad = -\frac{1}{2} \int_{q,{\rm shell}} \operatorname{tr} \left[ M_{BF} M^{-1}_{FF} M_{FB} M^{-1}_{BB} \right] + \cdots \nonumber\\
    &\quad = g_s^2 C_2\int_{q,{\rm shell}} \int_{x,y} \int_{q_1, q_2, q_3} \operatorname{tr} \left[ \bar{\psi}(x) \gamma_\mu e^{-i(q - q_1) \cdot x} S(q_1)\delta_{q_1 q_2} \gamma_\nu \psi(y) e^{-i (q_2 - q_3)\cdot y} D_{\mu\nu}(q_3) \delta_{q_3 q} \right] + \cdots \nonumber\\
    &\quad = g_s^2 C_2\int_{q,{\rm shell}} \int_{x,y} \int_{q_2} \operatorname{tr} \left[ \bar{\psi}(x) \gamma_\mu e^{-i(q - q_2) \cdot x} S(q_2) \gamma_\nu \psi(y) e^{-i (q_2 - q)\cdot y} D_{\mu\nu}(q) \right] + \cdots \nonumber\\
    &\quad = g_s^2 C_2\int_{q,{\rm shell}} \int_{x,y} \int_{p} \operatorname{tr} \left[ \bar{\psi}(x) \gamma_\mu S(q + p) \gamma_\nu \psi(y) D_{\mu\nu}(q) \right] e^{-i p \cdot (x - y)} + \cdots,
    \label{eq:generalTwoPointFlow}
\end{align}
where in the last line, we have defined the shifted momentum $p \equiv q_2-q$.
For the case with finite quark chemical potential, the flow is equivalent to the original flow with the replacement of $S(q_2) \rightarrow S(q_2^{\pm})$ and averaging them, with only the real part remaining.
Diagrammatically, this loop integral corresponds to the sunset diagram, which is shown in \Cref{fig:SunsetOfTwoPoint}.

We then project the flow onto the tensorial structure of the spatial Dirac operator $i\bvec{\gamma} \cdot \bvec{p}$ as we discussed in \Cref{sec:flowOfVkMainText}, we obtain
\begin{align}
    |\bvec{p}|^2 \partial_t \Gamma^{(2)}_{\bar{\psi}\psi, i\bvec{\gamma} \cdot \bvec{p}}(p) = g_s^2 C_2\int_{q,\rm shell} \operatorname{tr} \left[ \gamma_\mu S(p_\psi^\pm + q_\psi) \gamma_\nu \gamma_i \right] p_i D_{\mu\nu}(q_A)/ \mathcal{N},
\end{align}
where $\mathcal{N} = \operatorname{tr}\left[(\bvec{\gamma} \cdot \bvec{p})^2/|\bvec{p}|^2\right] = 4$.The normalized trace is obtained as 
\begin{align}
    &\operatorname{tr} \left[ \gamma_\mu S(p_\psi^\pm + q_\psi) \gamma_\nu \gamma_i \right] p_i D_{\mu\nu}(q_A) / \mathcal{N} \nonumber\\
    &\quad= \frac{(1-\xi)\frac{2\left( q_A \cdot(p_\psi^\pm+q_A) \right) \bvec{p}\cdot\bvec{q}}{(\omega_{A,n}^2 + |\bvec{q}|^2)} + (1+\xi) \bvec{p} \cdot (\bvec{p}+\bvec{q})}{(\omega_{A,n}^2 + |\bvec{q}|^2)\left[ \omega_{\psi,n}^{\pm,2} + (\bvec{q} + \bvec{p})^2 + M_\psi^2 \right]} \nonumber\\
    &\quad= \frac{2(1+\xi)\left( \omega_{A,n}\omega_{\psi,0}^\pm + \bvec{q}\cdot\bvec{p} + \omega_{A,n}^2 + |\bvec{q}|^2 \right)\bvec{q}\cdot\bvec{p} + (1+\xi)(\omega_{A,n}^2 + |\bvec{q}|^2) \left( \bvec{q}\cdot\bvec{p} + |\bvec{p}|^2 \right)}{(\omega_{A,n}^2 + |\bvec{q}|^2)^2 \left[ \omega_{\psi,n}^{\pm,2} + (\bvec{q} + \bvec{p})^2 + M_\psi^2 \right]} \nonumber\\
    &\quad= \frac{(3-\xi)(\omega_{A,n}^2 + |\bvec{q}|^2) \bvec{q}\cdot\bvec{p} + (1+\xi)(\omega_{A,n}^2 + |\bvec{q}|^2) |\bvec{p}|^2 + 2(1-\xi)(\omega_{A,n}\omega_{\psi,0}^\pm+\bvec{q}\cdot\bvec{p})\bvec{q}\cdot\bvec{p}}{(\omega_{A,n}^2 + |\bvec{q}|^2)^2 \left[ \omega_{\psi,n}^{\pm,2} + (\bvec{q} + \bvec{p})^2 + M_\psi^2 \right]}.
    \label{eq:ProjectingTwoPointFlow}
\end{align}
Then, we perform the derivative (momentum) expansion to the external spatial momentum $\bvec{p}$ in the denominator of \cref{eq:ProjectingTwoPointFlow} up to $\mathcal{O}(\bvec{p}^2)$ as
\begin{align}
    \text{\cref{eq:ProjectingTwoPointFlow}} &= \frac{(3-\xi)(\omega_{A,n}^2 + |\bvec{q}|^2) \bvec{q}\cdot\bvec{p} + (1+\xi)(\omega_{A,n}^2 + |\bvec{q}|^2) |\bvec{p}|^2 + 2(1-\xi)(\omega_{A,n}\omega_{\psi,0}^\pm+\bvec{q}\cdot\bvec{p})\bvec{q}\cdot\bvec{p}}{(\omega_{A,n}^2 + |\bvec{q}|^2)^2 \left[ \omega_{\psi,n}^{\pm,2} + |\bvec{q}|^2 + M_\psi^2 \right]} \nonumber\\
    &\quad \times \left[ 1 - \frac{2 \bvec{q} \cdot \bvec{p}}{\left[ \omega_{\psi,n}^{\pm,2} + |\bvec{q}|^2 + M_\psi^2 \right]} + \frac{4(\bvec{q}\cdot\bvec{p})^2}{\left[ \omega_{\psi,n}^{\pm,2} + |\bvec{q}|^2 + M_\psi^2 \right]^2} - \frac{|\bvec{p}|^2}{\left[ \omega_{\psi,n}^{\pm,2} + |\bvec{q}|^2 + M_\psi^2 \right]} \right] + \mathcal{O}(|\bvec{p}|^3).
    \label{eq:derivativeExpansion}
\end{align}

The momentum shell integral reads
\begin{align}
    \int_{q,\rm shell} = T\sum_{n=-\infty}^{\infty} \int\frac{\df^3 q}{(2\pi)^3} \, k\delta(|\bvec{q}| - k),
\end{align}
with
\begin{align}
    \int\df^3 q = \int 4\pi |\bvec{q}|^2 \, \df |\bvec{q}| \int_0^\pi \frac{\sin\theta}{2} \, \df\theta \equiv \int 4\pi |\bvec{q}|^2 \, \df |\bvec{q}| \int \df \Omega_3.
\end{align}
With the following identities
\begin{align}
    \int \df \Omega_3 = 1, \qquad \int \df \Omega_3 \, \cos \theta = 0, \qquad \int \df \Omega_3 \cos^2 \theta = \frac{1}{3}, \qquad \int \df \Omega_3 \cos^4 \theta = \frac{1}{5},
\end{align}
we obtain the flow after performing the solid angle integral
\begin{align}
    \int \df \Omega_3 \, \text{\cref{eq:derivativeExpansion}} &= \frac{1}{(\omega_{A,n}^2 + |\bvec{q}|^2)^2 \left[ \omega_{\psi,n}^{\pm,2} + |\bvec{q}|^2 + M_\psi^2 \right]} \nonumber\\
    &\quad \times \biggl[ 0 + (1+\xi)(\omega_{A,n}^2 + |\bvec{q}|^2) |\bvec{p}|^2 + \frac{2}{3}(1-\xi) |\bvec{q}|^2 |\bvec{p}|^2 \nonumber\\
    &\quad\quad\quad - \frac{\frac{2}{3}(3-\xi)(\omega_{A,n}^2 + |\bvec{q}|^2)|\bvec{q}|^2 |\bvec{p}|^2 + \frac{4}{3}(1-\xi)\omega_{A,n}\omega_{\psi,0}^\pm |\bvec{q}|^2 |\bvec{p}|^2}{\omega_{\psi,n}^{\pm,2} + |\bvec{q}|^2 + M_\psi^2} \nonumber\\
    &\quad\quad\quad + \frac{\frac{4}{3}(1+\xi)(\omega_{A,n}^2 + |\bvec{q}|^2)|\bvec{q}|^2 + \frac{8}{5}(1-\xi)|\bvec{q}|^4}{\left[ \omega_{\psi,n}^{\pm,2} + |\bvec{q}|^2 + M_\psi^2 \right]^2}|\bvec{p}|^4 \nonumber\\
    &\quad\quad\quad -\frac{(1+\xi)(\omega_{A,n}^2 + |\bvec{q}|^2) + \frac{2}{3}(1-\xi)|\bvec{q}|^2}{\omega_{\psi,n}^{\pm,2} + |\bvec{q}|^2 + M_\psi^2} |\bvec{p}|^4 \biggl] + \mathcal{O}(|\bvec{p}|^3).
\end{align}
Collecting the term proportional to $|\bvec{p}|^2$, and performing the momentum shell integral, we obtain the quark anomalous dimension within the current truncation scheme
\begin{align}
    \eta_\psi &= -g_s^2 C_2 T\sum_{n=-\infty}^{\infty} \int \frac{\df |\bvec{q}|}{2 \pi^2} \, |\bvec{q}|^2 k \delta(|\bvec{q}| - k) \biggl[ + \frac{(1+\xi)}{(\omega_{A,n}^2 + |\bvec{q}|^2) \left[ \omega_{\psi,n}^{\pm,2} + |\bvec{q}|^2 + M_\psi^2 \right]} + \frac{\frac{2}{3}(1-\xi)|\bvec{q}|^2}{(\omega_{A,n}^2 + |\bvec{q}|^2)^2 \left[ \omega_{\psi,n}^{\pm,2} + |\bvec{q}|^2 + M_\psi^2 \right]} \nonumber\\
    &\hspace{0.34\textwidth} - \frac{\frac{2}{3}(3 - \xi) |\bvec{q}|^2 \omega_{A,n} \omega_{\psi,0}^\pm}{(\omega_{A,n}^2 + |\bvec{q}|^2) \left[ \omega_{\psi,n}^{\pm,2} + |\bvec{q}|^2 + M_\psi^2 \right]^2} - \frac{\frac{4}{3}(1 - \xi) |\bvec{q}|^2}{(\omega_{A,n}^2 + |\bvec{q}|^2)^2 \left[ \omega_{\psi,n}^{\pm,2} + |\bvec{q}|^2 + M_\psi^2 \right]^2} \biggl] \nonumber\\
    &= -\frac{g_s^2 C_2}{2\pi^2}\frac{T}{k} \sum_{n=-\infty}^{\infty} \biggl[ \frac{(1+\xi)}{(\tilde{\omega}_{A,n}^2 + 1) \left[ \tilde{\omega}_{\psi,n}^{\pm,2} + 1 + \tilde{M}_\psi^2 \right]} + \frac{\frac{2}{3}(1-\xi)}{(\tilde{\omega}_{A,n}^2 + 1)^2 \left[ \tilde{\omega}_{\psi,n}^{\pm,2} + 1 + \tilde{M}_\psi^2 \right]} - \frac{\frac{2}{3}(1-\xi)}{(\tilde{\omega}_{A,n}^2 + 1) \left[ \tilde{\omega}_{\psi,n}^{\pm,2} + 1 + \tilde{M}_\psi^2 \right]^2} \biggl] \nonumber\\
    &= -\frac{g_s^2 C_2}{2\pi^2} \biggl[ (1+\xi) \mathcal{FB}_{(1,1)}(M_\psi^2,0,T,\pm \mu_q,\pi T) + \frac{2}{3}(1-\xi) \mathcal{FB}_{(1,2)}(M_\psi^2,0,T,\pm \mu_q,\pi T) \nonumber\\
    &\hspace{0.1\textwidth} -  \frac{2}{3}(1-\xi) \mathcal{FB}_{(2,1)}(M_\psi^2,0,T,\pm \mu_q,\pi T) \biggl].
    \label{eq:quarkAnomalousDimension}
\end{align}
In the last line of \cref{eq:quarkAnomalousDimension}, we utilize the same notation of the threshold functions defined in Ref.~\cite{Fu:2019hdw}, that
\begin{align}
    \mathcal{FB}_{(n_f,n_b)}(m_f^2,m_b^2,T,+\mu_q,p_0) \equiv \frac{T}{k}\sum_{n=-\infty}^{\infty}\biggl( \frac{1}{\tilde{\omega}_{\psi,n}^{+,2} + 1 + \tilde{m}_f^2} \biggl)^{n_f}\biggl( \frac{1}{\tilde{\omega}_{A,n}^{+,2} + 1 + \tilde{m}_b^2} \biggl)^{n_b},
\end{align}
where those terms with a tilde denote the dimensionless quantities, e.g., $\tilde{m} = m/k$.
At the lowest order, the threshold function is also shown in the Ref.~\cite{Fu:2019hdw}, and a higher order is found by taking the derivative with respect to the mass
\begin{align}
    &\mathcal{FB}_{(n_f+1,n_b)}(m_f^2,m_b^2,T,+\mu_q,p_0) = -\frac{1}{n_f} \partial_{m_f^2}\mathcal{FB}_{(n_f,n_b)}(m_f^2,m_b^2,T,+\mu_q,p_0), \nn
    &\mathcal{FB}_{(n_f,n_b+1)}(m_f^2,m_b^2,T,+\mu_q,p_0) = -\frac{1}{n_b} \partial_{m_b^2}\mathcal{FB}_{(n_f,n_b)}(m_f^2,m_b^2,T,+\mu_q,p_0).
\end{align}
Notice that after taking the average of $\pm\mu_q$ as we discussed below \cref{eq:generalTwoPointFlow}, the result is equivalent to taking one of $\pm\mu_q$ and taking the real part, thus the final expression of the quark anomalous dimension reads
\begin{align}
    \eta_\psi &= -\frac{g_s^2 C_2}{2\pi^2} \operatorname{Re}\biggl[ (1+\xi) \mathcal{FB}_{(1,1)}(M_\psi^2,0,T,\mu_q,\pi T) + \frac{2}{3}(1-\xi) \mathcal{FB}_{(1,2)}(M_\psi^2,0,T,\mu_q,\pi T) \nonumber\\
    &\hspace{0.15\textwidth} -  \frac{2}{3}(1-\xi) \mathcal{FB}_{(2,1)}(M_\psi^2,0,T,\mu_q,\pi T) \biggl].
    \label{eq:quarkAnomalousDimensionReal}
\end{align}

\twocolumngrid
%%%%%%%%%%%%%%%%%%%%%%%%%%%%%%%%%%%%%%%%%%%%%%%%%%%%%%%%%%%%%%%%%%%%%%
%\bibliographystyle{JHEP} 
\bibliography{refs}
\end{document}